\documentclass[a4wide,11pt]{article}
\usepackage{fullpage}
\usepackage[english]{babel}
\usepackage{amsmath,amsthm}
\usepackage{amsfonts}
\usepackage{graphicx}
\usepackage{amssymb}

\RequirePackage{natbib}
\RequirePackage[OT1]{fontenc}
\begin{document}
\begin{titlepage}
\begin{center}
\vspace*{1cm}
\Huge
{\bf Model based functional clustering of varved lake sediments}

\vspace{2cm}
\huge
{Per Arnqvist \& Sara Sj\"{o}stedt de Luna }

\vfill%
\end{center}
\begin{abstract}
In this paper we propose a model-based method for clustering subjects for which functional data together with covariates are observed. The model allows the covariance structures within the different clusters to be different. The model thus extends a model proposed by James and Sugar (2003). We derive an EM algorithm to estimate the parameters. The method is applied to annually laminated (varved) sediment from lake Kassj\"on in northern Sweden, to infer on past climate changes.

\date{\today}
\end{abstract}
\end{titlepage}
\newpage\null\thispagestyle{empty}
\newpage
\setcounter{page}{1}
\section{Introduction}

There is an increasing literature addressing how to cluster functional data (curves), ranging from non-parametric methods \citep[see, e.g.][]{abraham, garcia, tarpey, serban} to model-based methods \citep[see e.g.][]{james, luan, chiou}. In this paper we propose a model-based method to cluster independent subjects for which functional data as well as covariates are observed. The aim is to cluster the subjects into homogenous groups taking into account both the functional data and the covariates. In model-based clustering it is assumed that the observations are generated according to a mixture distribution with G components (clusters). The approach taken here extend
James and Sugar's (2003) proposed model-based functional clustering method for sparsely distributed functional data. The (discretely) observed random functions  are there assumed to be Gaussian with a mean structure that depends on the cluster but with the same covariance structures for all clusters. In this paper, the model-based functional clustering method of James and Sugar is extended to allow for different covariance structures within the different clusters including additional covariates.  We propose an EM algorithm to estimate the parameters of the model.

Our model-based functional clustering method is motivated by and applied to varved (annually laminated) sediment from lake Kassj\"on in northern Sweden, aiming to infer on past climate changes. The varved sediment of lake Kassj\"on covers approximately 6400 years. The varves (years) are clustered into similar groups based on their seasonal patterns (functional data) and additional covariates,  all potentially carrying information on past climate/weather (cf. Section 4 for more details). The time dynamics of the resulting clusters are then used to infer on past climate.
Functional clustering has been applied to the seasonal patterns of the sediment data of Kassj{\"o}n before, \citep[see][]{arnqvist00,abramowicz}, but this is the first time that both the seasonal patterns and additional covariates are used when clustering the varves.

The paper is structured as follows. The model-based functional clustering models with and without covariates are described in Section 2 allowing for different covariance structures in different clusters. In Section 3  ways of determining the number of clusters are discussed. Section 4 applies the model-based functional clustering methods  to annually laminated sediment of lake Kassj\"on. Concluding remarks are given in Section 5. In Appendix A the EM-algorithms for the clustering models of Section 2 are derived, and in Appendix B, some implementation details of the EM algorithm are discussed.

\section{The clustering model}
\subsection{Model for functional data}
Assume that for a set of $N$ independent subjects, we have observed a random function for each subject, and are interested in clustering the subjects into $G$ (homogenous) groups. In this paper, we adopt a model-based approach to clustering the subjects. The model is described below. For each subject $i$ we observe the true continuous random function $g_i(t)\in D$ with measurement error, over a set of $n_i$ time points $t_{i1},...,t_{in_i}$, and thus observe
\[
y_i(t_{ij}) = g_i(t_{ij})+\epsilon_i(t_{ij}), \hspace{3mm} j=1,...,n_,\hspace{2mm} i=1,...,N
\]
where $y_i(t)$ is the observered function and $\epsilon_i(t)$ assumed to be independent and identically distributed (iid) measurement errors, normally distributed with mean zero and variance $\sigma^2$. Let $\boldsymbol{y}_i$, $\boldsymbol{g}_i$ and $\boldsymbol{\epsilon}_i$ be the corresponding $n_i$-dimensional vectors for subject $i$, corresponding to the observed values, true values and measurment error, respectively. We further assume that the smooth function $g_i(t)$ can be expressed as
\[
g_i(t) = \boldsymbol{\phi}(t)^T \boldsymbol{\eta}_i,
\]
where $\boldsymbol{\phi}(t)=[\boldsymbol{\phi}_1(t);,...,\boldsymbol{\phi}_p(t)]^T$ is a $p$-dimensional vector of known basis functions evaluated at time t, e.g. B-splines, Fourier or Wavelet basis, see \citep{ruppert} and $\boldsymbol{\eta}_i$ a vector of unknown (random) coefficients. The $\boldsymbol{\eta}_i$'s are modelled as
\[
\boldsymbol{\eta}_i = \boldsymbol{\mu}_{z_i} + \boldsymbol{\gamma}_i \in N_p(\boldsymbol{\mu}_{z_i},\Gamma_{z_i}),
\]
where $\boldsymbol{\mu}_{z_i}$ are a vector of expected spline coefficients for a cluster and $z_i$ denotes the unknown cluster membership, with $P(z_i=k)=\pi_k$,  $k=1,...,G$. The random variable $\boldsymbol{\gamma}_i$ corresponds to subject-specific within-cluster variability. Note that this variability is allowed to be different in different clusters, due to $\Gamma_{z_i}$. If desirable, given that subject $i$ belongs to cluster $z_i=k$, a further parametrization of $\boldsymbol{\mu}_{k}, k=1,...,G$ may prove useful, for producing low-dimensional representations of the curves as suggested by \cite{james}:
\[
\boldsymbol{\mu}_{k} = \boldsymbol{\lambda}_{0}+ \Lambda \boldsymbol{\alpha}_{k}
\]
where $\boldsymbol{\lambda}_{0}$ and $\boldsymbol{\alpha}_{k}$ are $p$- and $h$-dimensional vectors respectively and $\Lambda$ is a $p \times h$ matrix with $h \leq G-1$. Choosing $h<G-1$ may be valuable, especially for sparse data, cf \citep{james}. In order to ensure identifiability, some restrictions need to be put on the parameters. Imposing the restriction that $\sum_{k=1}^G \boldsymbol{\alpha}_k=\boldsymbol{0}$ implies that $\boldsymbol{\phi}^T(t) \boldsymbol{\lambda}_0$ can be viewed as the overall mean curve. Depending on the choice of $h,p$ and $G$, further restrictions may need to be imposed in order to have identifiability of the parameters ($\boldsymbol{\lambda}_{0}, \Gamma$ and $\boldsymbol{\alpha}_{k}, k=1,...,G$ are confounded if no restrictions are imposed), see \citep{james} for some suggestions.
In vector-notation we thus have
\[
\boldsymbol{y}_i = {\phi}_i (\boldsymbol{\lambda}_0 +  \Lambda \boldsymbol{\alpha}_{z_i}+\boldsymbol{\gamma}_i)+\boldsymbol{\epsilon}_i, i=1,...,N,
\]
where $\phi_i=\left[\boldsymbol{\phi}^T(t_{i1}); ...; \boldsymbol{\phi}^T(t_{in_i})\right]$ is an $n_i \times p$ matrix and $\boldsymbol{\epsilon}_i \in N_{n_i}(0,R)$ where $R=\sigma^2 \mathbb{I}_{n_i}$. We will also assume that the $\boldsymbol{\gamma}_i$'s, $\boldsymbol{\epsilon}_i$'s and the $z_i$'s are independent. Hence, given that subject $i$ belongs to cluster $z_i=k$ we have
\begin{equation}
\boldsymbol{y}_i | z_i=k \in N_{n_i}({\phi}_i (\boldsymbol{\lambda}_0 +  \Lambda \boldsymbol{\alpha}_{k}),\phi_i \Gamma_k \phi_i^T+R).\label{eqn1}
\end{equation}
Based on the observed data $\boldsymbol{y}_1,...,\boldsymbol{y}_N$, the parameters $\theta$
of the model can be estimated by maximizing the observed likelihood
\begin{equation}
L_o(\theta|\boldsymbol{y}_1,...,\boldsymbol{y}_N)=\prod_{i=1}^N \sum_{k=1}^G \pi_k f_k(\boldsymbol{y}_i,\theta),\label{eqn2}
\end{equation}
where $\theta = (\boldsymbol{\lambda}_0,\Lambda,\boldsymbol{\alpha}_1,...,\boldsymbol{\alpha}_G,\pi_1,...,\pi_G,
\sigma^2,\Gamma_1,...,\Gamma_G),$ and $f_k(\boldsymbol{y}_i,\theta)$ is the normal density given in (\ref{eqn1}). Note that throughout the paper $\theta$ will denote all scalar, vectors and matrices of parameters to be estimated.

In line with \cite{james}, we propose to maximize (\ref{eqn2}) by an EM-algorithm \citep{dempster}. The EM-algorithm iteratively performs two steps, an expectation and a maximization step. First, the expected value of the complete likelihood given the observed data and starting values for the parameter $\theta$ is computed. Then this expected value is maximized with respect to $\theta$. The updated parameter estimates are plugged into the expectation and a new iteration begins. The algorithm ends when the parameter changes between iterations are sufficiently small. In our situation, the complete likelihood is chosen to be
\[
L_c(\theta) = \prod_{i=1}^N f(\boldsymbol{y}_i, \boldsymbol{\gamma}_i, z_i | \theta).
\]
Note that the joint density $f(\boldsymbol{y}_i, \boldsymbol{\gamma}_i, z_i | \theta)$ can be factorized as
\[
f(\boldsymbol{y}_i, \boldsymbol{\gamma}_i, z_i | \theta) = f(\boldsymbol{y}_i | \boldsymbol{\gamma}_i, z_i) f(\boldsymbol{\gamma}_i| z_i) f(z_i).
\]
We have that $\boldsymbol{y}_i |\boldsymbol{\gamma}_i, z_i=k$ is normally distributed with
\[
\boldsymbol{y}_i |\boldsymbol{\gamma}_i, z_i=k \in N_{n_i}(\phi_i(\boldsymbol{\lambda}_0+\Lambda \boldsymbol{\alpha}_k+\boldsymbol{\gamma}_i),R),
\]
and
\[
\boldsymbol{\gamma}_i|z_i=k \in N_p(0,\Gamma_k).
\]
For notational convenience, let us denote $\boldsymbol{z}_i=(z_{i1},...,z_{iG})$, where $z_{ik}=1$ if $z_{i}=k$ and 0 otherwise. Then $\boldsymbol{z}_i$ follows a multinomial distribution with
\[
f(\boldsymbol{z}_i)=\pi_1^{z_{i1}}...\pi_G^{z_{iG}}.
\]
Under the assumption of independence between subjects, we thus have that the complete log likelihood, up to an additive constant, is
\begin{align}
l(\theta) &  =\sum
_{i=1}^{N}\sum_{k=1}^{G}z_{ik}\log(\pi_{k})\nonumber\\
&  -\frac{1}{2}\sum_{i=1}^{N}\sum_{k=1}^{G}z_{ik}\left[  \log\left\vert \Gamma_{k}\right\vert
+\boldsymbol{\gamma}_{i}^{T}\Gamma_{k}^{-1}\boldsymbol{\gamma}_{i}\right]\nonumber\\
&  -\frac{1}{2}\sum_{i=1}^{N}\sum_{k=1}^{G}z_{ik}\left[  n_{i}\log\sigma
^{2}+\frac{1}{\sigma^{2}}\left\vert \left\vert \boldsymbol{y}_{i}-\phi_{i}%
(\boldsymbol{\lambda}_{0}+\Lambda\boldsymbol{\alpha}_{k}+\boldsymbol{\gamma}_{i})\right\vert \right\vert ^{2}\right].
\label{eqn3}%
\end{align}
The expected value of (\ref{eqn3}), given the data $(\boldsymbol{y}_1,...,\boldsymbol{y}_N)$ and current parameter estimates is then maximized, (see further details in Appendix A). The EM-algorithm proceeds in an iterative manner until the parameter estimates have converged.

Once the parameters have been estimated via the EM-algorithm, the cluster label assignment of subject $i$ will rely on the posterior probabilities
\[
\hat{\pi}_{k|i} = P(z_i=k|\boldsymbol{y}_i,\theta) =\frac{f_k(\boldsymbol{y}_i,\theta)\pi_k}{\sum_{k=1}^G f_j(\boldsymbol{y}_i,\theta)\pi_j},\hspace{2mm} k=1,...,G,
\]
where $\theta$ is replaced by the estimator $\hat\theta$. Here, subject $i$ is assigned to the cluster whose label correspond to the largest posterior probability.
\subsection{Model for functional data and covariates}
If additional covariates  have been observed for each subject besides the functional data, they can also be included in the model when clustering the subjects. 
In this section we extend the functional clustering model, described in the previous section, to also include covariates. Given that individual $i$ belongs to cluster $k, (z_{i}=k)$ the $r$ covariates $\boldsymbol{x}_i \in \mathbb{R}^r$ are assumed to have mean value $\boldsymbol{\upsilon}_k$ and moreover $\boldsymbol{x}_{i} = \boldsymbol{\upsilon}_{k} + \boldsymbol{\delta}_{i} + \boldsymbol{e}_i,$
where we assume that $\boldsymbol{\delta}_{i}|z_{i}=k \sim N(\boldsymbol{0}, D_k)$ is the random deviation within cluster and $\boldsymbol{e}_i \sim N(\boldsymbol{0},\sigma_x^2 \mathbb{I}_r)$ independent remaining unexplained variability. Denote the observed data for subject $i$ by
\[
\boldsymbol{u}_i = \left( \begin{array}{c} \boldsymbol{y}_{i}\\ \boldsymbol{x}_{i} \end{array}\right ) \in \mathbb{R}^{n_i + r},
\]
and let $\boldsymbol{\xi}_{i}= (\boldsymbol{\gamma}_{i}^T,\boldsymbol{\delta}_{i}^T)^T$ with $\boldsymbol{\xi}_{i}|z_{i}=k \sim N_{n_i + r}(\boldsymbol{0},\Delta_k)$, where
\[
\Delta_k = \left( \begin{array}{cc} \Gamma_k & L_k\\ L_k^T & D_k \end{array}\right ).
\]
\begin{flushleft}
Given that subject $i$ belongs to cluster $k$, we can then write
\end{flushleft}
\[
\boldsymbol{u}_i = S_i (\boldsymbol{\mu}_k + \boldsymbol{\xi}_i) + \boldsymbol{\zeta}_i,
\]
where
\[
S_{i} = \left[ \begin{array}{cc}
\phi_i & \boldsymbol{0} \\
\boldsymbol{0} & \mathbb{I}_r
\end{array}
\right]
,
\boldsymbol{\mu}_{k} = \left( \begin{array}{c}
\boldsymbol{\lambda}_{0}+\Lambda \boldsymbol{\boldsymbol{\alpha}}_{k} \\
\boldsymbol{\upsilon}_{k}\\
\end{array}
\right),
\]
and
$\boldsymbol{\zeta}_i =\left( \epsilon_i^T, e_i^T\right)^T\sim N_{n_i+r}(\boldsymbol{0},R)$, with
\[
R = \left[
\begin{array}{cc}
\sigma^2 \mathbb{I}_{n_i} & \large{\boldsymbol{0}}\\
\large{\boldsymbol{0}} & \sigma^2_{x} \mathbb{I}_{r}\\
\end{array}
\right],
\]
and hence,
\begin{align}
\boldsymbol{u}_i|z_i=k \sim N_{n_i+r}(S_i \boldsymbol{\mu}_k,S_i\Delta_k S_i^T+R).\label{fu}
\end{align}
Note that this model incorporates the dependence between covariates and the random functions via the random coefficients of the basis functions.
The unknown vector of parameters to be estimated is\\
 $\theta = (\boldsymbol{\pi},\Delta_{1},...,\Delta_G,\sigma^{2},\sigma^2_x,
 \boldsymbol{\upsilon}_{1},..., \boldsymbol{\upsilon}_{G}, \boldsymbol{\lambda}_{0},\Lambda,{\boldsymbol{\alpha}}_1,...,{\boldsymbol{\alpha}}_G)$.
Noting that the complete likelihood for subject $i$ can be factorized as
\begin{align}
f(\boldsymbol{u}_i,\boldsymbol{\xi}_i,\boldsymbol{z}_i)=
f(\boldsymbol{y}_i|\boldsymbol{\xi}_i,\boldsymbol{z}_i)
f(\boldsymbol{x}_i|\boldsymbol{\xi}_i,\boldsymbol{z}_i)
f(\boldsymbol{\xi}_i|\boldsymbol{z}_i)f(\boldsymbol{z}_i),
\label{app.fact}
\end{align}
and by the independence between individuals the complete log likelihood, up to an additive constant, is given by
\begin{align}
l(\theta) &  \propto \sum
_{i=1}^{N}\sum_{k=1}^{G}z_{ik}\log(\pi_{k}) \nonumber \\
&  -\frac{1}{2}\sum_{i=1}^{N}\sum_{k=1}^{G}z_{ik}\left[  \log\left\vert \Delta_{k}\right\vert
+\boldsymbol{\xi}_{i}^{T}\Delta_{k}^{-1}\boldsymbol{\xi}_{i}\right] \nonumber \\
&  -\frac{1}{2}\sum_{i=1}^{N}\sum_{k=1}^{G}z_{ik}\left[  n_{i}\log\sigma
^{2}+ \frac{1}{\sigma^{2}}\left\vert \left\vert \boldsymbol{y}_{i}-\phi_{i}
(\boldsymbol{\lambda}_{0}+\Lambda\boldsymbol{\alpha}_{k}+\boldsymbol{\gamma}_{i})\right\vert \right\vert ^{2}
\right] \nonumber\\
& -\frac{1}{2}\sum_{i=1}^{N}\sum_{k=1}^{G}z_{ik} \left[  r \log(\sigma_x^2) + \frac{1}{\sigma_x^2} ||\boldsymbol{x}_{i} - \boldsymbol{\upsilon}_{k}+\boldsymbol{\delta}_{i}||^2 \right].
\label{app.single.sigma}
\end{align}

\begin{flushleft}
Using the EM-algorithm we want to maximize the expected value of $l(\theta)$ given the observed data $\boldsymbol{U} = (\boldsymbol{u}_1, ...., \boldsymbol{u}_N)$ and starting parameter values
$
\theta^{[0]} 
$
i.e. maximize $E[l(\theta)|\boldsymbol{U}, \theta^{[0]}]$, with respect to $\theta$. More computational details for the EM-algorithm 
is found in Appendix A.
Again, once the parameter estimates $\hat{\theta}$ have been determined, cluster label assignment for each subject is determined by the maximum of the posterior probabilities
\begin{align}
P(z_i=k|\boldsymbol{u}_i,\hat{\theta})=\frac{f_k(\boldsymbol{u}_i,\hat{\theta}) \hat{\tilde{\pi}}_k}{\sum_{j=1}^G f_j(\boldsymbol{u}_i,\hat{\theta}) \hat{\tilde{\pi}}_j}, \hspace{2mm} k=1,...,G, \label{maxpost}
\end{align}
where $f_k(\boldsymbol{u}_i,\theta)$ corresponds to the distribution in (\ref{fu}).
\end{flushleft}
\section{Determining the number of clusters}
So far we have assumed that the number of clusters $G$ is known. However in practice it needs to be determined from the data. Several proposals have been suggested in the literature, \citep{abraham, garcia, tarpey, serban}.
In a model-driven approach it is natural to find the number of clusters needed based on the observed log likelihood, e.g. through information criteria such as AIC or BIC or by studying (relative) changes of the observed log likelihood as the number of clusters is increased.
The information criteria are calculated as
AIC = $2 \cdot m - 2 \log(L_o(\hat{\theta}))$ and BIC = $m \cdot \log(N) -2  \log(L_o(\hat{\theta}))$, where $m$ is the number of parameters and $\log(L_o(\hat{\theta}))$ is the observed log likelihood of the data consisting of $N$ subjects. For large number of observations the penalization of the log likelihood with respect to the number of parameters $m$ in AIC and BIC is often of minor importance. The best model according to AIC or BIC is the one that minimizes AIC or BIC.
Another alternative could be to chose the number of clusters where the relative change in the observed log likelihood
\begin{align}
\frac{\log L_o(\hat{\theta}|G+1)-\log L_o(\hat{\theta}|G)}{\log L_o(\hat{\theta}|G+1)},
\label{relativelogL}
\end{align}
levels out and is small. The relative differences for AIC, BIC or some parameter estimates like $\hat{\sigma}^2$ or $(\hat{\sigma}^2+\hat{\sigma}_x^2)$ could also be used for this purpose. Yet, another tool suggested by \citet{james} is the ``distortion function''
\begin{align}
d_G = \frac{1}{p}\min_{c_1, \dots, c_G} E[\boldsymbol{\eta}_i -\boldsymbol{c}_{z_i}]^T \Gamma^{-1}_{z_i} E[\boldsymbol{\eta}_i -\boldsymbol{c}_{z_i}],\label{dK}
\end{align}
where the $\boldsymbol{\eta}_i$'s are the random spline coefficients in the clustering model, $p$ is the number of spline basis, and $\boldsymbol{c}_{1},...,\boldsymbol{c}_{G} $ are the $G$ cluster (coefficient) medoids. The distortion, $d_G$, is the average Mahalanobis distance between each $\boldsymbol{\eta}_i$, and its closest cluster medoid (coefficient) $\boldsymbol{c}_{z_i}$. The difference
\begin{align}
\Delta d^{-b}_{G,G-1}=d^{-b}_{G}-d^{-b}_{G-1}\label{delta},
\end{align}
is then plotted for different values of G, where $b$ is a parameter that needs to be determined. See Appendix C for computational details.
\section{Analyzing the varved sediment of lake Kassj\"on}
We will now analyze the annually laminated (varved) sediment of lake Kassj\"on, situated in northern Sweden. Varved lake sediment has the potential to play an important role for understanding past climate with their inherent annual time resolution and within-year seasonal patterns, see part of the varved sediment of lake Kassj\"on in Figure 1. The varved patterns of lake Kassj\"on have the following origin. During spring, in connection to snow melt and spring runoff, minerogenic material is transported from the catchment area into the lake through four small streams, which gives rise to a bright colored layer, (large gray-scale values) \citep{petterson1}. During summer, autochthonous organic matter sinks to the bottom and creates a darker layer (lower gray-scale values). Finally, during winter, when the lake is ice-covered, fine organic material is deposited, resulting in a thin blackish winter layer (lowest gray-scale values). Figure 1 reveals substantial within- and between year variation, reflecting the balance between minerogenic and organic material. The properties of each varve reflect, to a large extent, weather conditions and internal biological processes in the lake the year that the varve was deposited. The minerogenic input reflects the intensity of the spring run-off, which is dependent on the amount of snow accumulated during the winter, and hence the variability in past winter climate. Note also that the seasonal patterns may indicate important weather information. For example, a pronounced spring peak may correspond to a winter with rich amounts of snow, and a low spring peak a winter with less snow. A substantial flatter part after the spring peak could correspond to a thick organic layer, perhaps indicating a warmer summer.

\subsection{Preliminaries}

The information in the varved sediment of lake Kassj\"on was registered by image analysis as gray-scale values \citep[cf.][]{petterson2,petterson3}. The raw data set consists of a series of averages of five-pixel slices subjectively chosen from representative parts of the varved sediment images, cf Figure 1, \citep{petterson2,petterson3}.
The data were recorded as gray-scale values with yearly deliminators giving 6388 years spanning over the time period 4386 B.C. until A.D. 1901. Of the 6388 varves 62 of them had no gray scale values recorded and were therefore treated as missing. See \cite{arnqvist00} for more information.
The data thus consists of $N=6326$ (subjects) years and the $n_{i}$ observations per year ranges from 4 to 37. For each year $i$ we observe the seasonal pattern in terms of average gray-scale values ($y_i$'s) of the five pixel slices at $n_i$ time points (pixels). 

In order to make the seasonal patterns comparable we first put them on the same time scale [0,1], such that pixel position $j$ at year $i$ corresponds to position $\tilde{t}_{ij}=(j-1)/(n_i-1), \hspace{2mm} j=1,...,n_i,\hspace{1mm} i=1,...,N.$
To make the patterns more comparable (with respect to weather/climate) they were further aligned by landmark registration, synchronizing the first spring peaks, that are directly related to the spring flood that occurs approximately the same time each year. More specifically, we used the first (spring) peak landmarks ($L_i$'s) identified by \cite{arnqvist00}, and then warped the time points according to \[
t_{ij} =w(\tilde{t}_{ij})= \begin{cases} \tilde{t}_{ij} b_i & \mbox{if } \tilde{t}_{ij}<L_i \\
(\tilde{t}_{ij}-1) d_i+1 & \mbox{if } \tilde{t}_{ij}\geq L_i \end{cases}
\]
 where $M_L=0.2944$, $b_i=M_L/L_i$ and $d_i=(1-M_L)/(1-L_i)$, $i=1,...,N$.

Focusing on the functional forms of the seasonal patterns we finally centered them within years and worked with (the centered values) $y_i(t_{ij})-\bar{y}_i, \hspace{2mm} j=1,...,n_i,\hspace{1mm} i=1,...,N$, where $\bar{y}_i = \sum_{j=1}^{n_i} y_i(t_{ij})/n_i$ is the mean grey scale value of varve (year) $i$. In addition to the seasonal patterns we also include 3 covariates; the mean grey scale $x_{1i}=\bar{y}_i$, the varve width (proportional to $n_i$) $x_{2i} = n_i$ and the minerogenic accumulation rate (mg/$cm^2$) corresponding to the accumulated amount of minerogenic material per $cm^2$ in varve $i$, $x_{3i}$, \citep[for details see][]{petterson1}.

\subsection{Clustering the varved sediment of lake Kassj\"on}
Based on the functional data and the three covariates we fit the clustering model described in Section 2.2 for $G=2,3,...,11$ clusters. After some preliminary investigation we decided to use $p=8$ cubic B-spline basis functions formed from 6 evenly distributed knots $\kappa_r, r=(r-1)/5$, as our basis functions $\boldsymbol{\phi}(t)$ (see Figure \ref{basis} for an illustration).
For each given number of clusters $G=2,3,..., 11$ we initiated the corresponding EM algorithm with starting values for the parameters (see Appendix B for proposals). For $G=2,..., 9$ we set $h=G-1$ and for $G=10, 11$, $h=8$ was chosen. Similar to \cite{james} we used small absolute relative changes in $\sigma^2+\sigma_x^2$ between two successive EM iterations as convergence criterion, i.e. when the absolute relative changes were smaller than $0.001$ the algorithm stopped. To decide the number of clusters, we studied the relative differences in a) the observed log likelihood (\ref{relativelogL}), b) $\sigma^2+\sigma_x^2$, c) AIC and BIC and finally, d) the difference in distortion functions (\ref{delta}) with b=4 (see Figure \ref{reldiff} for $G=2,...,11$). From this information we conclude that $G=7$ clusters seems feasible.

The estimated covariance matrices, $\Delta_k$, $k=1,...,7$, are given in Figure \ref{covar}. In the figure the spline coefficients are numbered 1--8 and the covariates numbered 9--11 in the order MinAR, Number of observations and Mean grey scale values, respectively. It reveals that there is substantially larger variability in the covariates compared to the spline coefficients. We also see that the $\Delta_k$:s varies substantially between clusters, especially for the covariates.
In Figure \ref{corell} the correlation matrices are given corresponding to Figure \ref{covar}. Here it is easier to reveal the dependency among and between the spline coefficients and the covariates. It can be seen that the correlation differs between different clusters, especially for the correlation between the spline coefficients and the covariates. The higher the signal in the covariates is then the higher the correlation becomes. It can also be noted that neighbouring spline coefficients are negatively correlated.

The posterior probabilities (\ref{maxpost}) were calculated for each varve and cluster labels given as the maximum value of the posterior probabilities.
For illustration, the posterior probabilities  are given for two years together with the estimated cluster mean curves, $\boldsymbol{\phi}(t)^T (\hat{\boldsymbol{\lambda}}_0+\hat{\Lambda} \hat{\boldsymbol{\alpha}}_k), k=1,...,7$  (solid colored lines), and estimated expected yearly curve given $\boldsymbol{u}_i$, i.e.  $\boldsymbol{\phi}(t)^T E[\boldsymbol{\eta}_i|\boldsymbol{u}_i,\hat{\theta}]$ (dashed black lines), see Figures \ref{covill} and \ref{covill2}.

Figures \ref{fourcluster} and \ref{threecluster} summarize the information in the achieved 7 clusters and their time dynamics. In each of the seven panels the overall mean curve is given, $\phi(t)^T \hat{\boldsymbol{\lambda}}_0$ (dashed black line) together with the cluster mean curve (red solid line). In the top right corner, within each cluster, the mean values of the three covariates for that cluster are given. Box plots for the covariates of each cluster are given in Figure \ref{boxplots}, upper panels.
The seasonal patterns of the 7 clusters are similar to those found by \cite{arnqvist00}. There are a few clusters with pronounced high spring peak, clusters 1, 3 and 7 (which probably correspond to winters with high snow accumulation); one rather flat profile, cluster 6 (probably corresponding to years with mild winters); a pronounced double peaked profile, cluster 2 (where the second peak might indicate a fall storm with heavy rain); and finally two medium spring peak patterns, clusters 4 and 5.
Note that the mean gray-scale values and MinAr tend to be high for clusters with pronounced high spring peak and low for cluster 6 with the flat pattern. Also note that the doubled peaked cluster 2 also has high mean gray-scale value and MinAR. Cluster with 2 peaks in general has thicker varves compared to the other clusters.

The time dynamics of each cluster are represented in Figures \ref{fourcluster} and \ref{threecluster} as the number of years in cluster $k$ over 50 year bins (black solid lines with colored dots).
The colored dots correspond to the average maximum posterior probabilities of the years included in that 50 year bin of that particular cluster, which gives an indication of the certainty of the cluster label assignment in each bin. Overall it can be seen that the (max) posterior probabilities of the assigned cluster labels often are very high. The median of all maximum posterior probabilities are 0.9523 and 132 years had a maximum posterior probability less than 0.5.
When it comes to the time dynamics and climate interpretation, cluster 6 can be useful. It can be seen that a very high peak occurred about 1200 BC. Next peak was around 500 BC. Then AD 500, AD 900 and AD 1100  also had some peaks with high frequencies of the flat profile. This is probably an indication of time periods with warmer winter conditions.


As a comparison we also clustered the varves solely based on the seasonal patterns (functional data).
By again using $p=8$ cubic B-splines basis functions the model in Section 2.1 was estimated for $G=2,...,11$ clusters and the corresponding quantities from Section 3 computed, as presented in Figure \ref{reldiff}, upper panels. Again $G=7$ clusters seem to be appropriate.
Time dynamics, seasonal profiles and mean values for each of the three covariates (note that they were not taken into account when clustering) are given for all 7 clusters in Figures \ref{fourcluster.nocov} and \ref{threecluster.nocov}. The general trends compared to clustering with covariates (Figures \ref{fourcluster} and \ref{threecluster}) are still the same but in general the average posterior probabilities in the 50-year bins are smaller than those obtained when clustering with covariates. For the model without covariates the median posterior probabilities becomes 0.9473 and 116 years less than 0.5. This indicates a higher stability/certainty when adding covariates in the clustering model. The mean cluster curves are pretty similar but we see a movement of years that changed clusters, of around 25\%, which means that the covariates affect the clustering.
It can also be seen that the levels of the mean grey scale values and MinAR vary more between clusters when using covariates in the clustering process.

\section{Concluding remarks}
We have proposed a model-based functional clustering method to group subjects where functional data and covariates are observed. We allow the covariance structure to be different in different clusters and suggest the EM-algorithm to estimate the parameters of the model.

When estimating the model based on the varved lake sediment data of Kassj\"on  significant differences in the covariance matrices of the different clusters are revealed, exemplifying the usefulness of allowing different covariance matrices. For smaller sample sizes it may not be sufficient to allow so many parameters to be estimated, thus a more parsimonious model might be better suited, for example, as in \cite{james}.
For the Kassj\"on data, clustering with covariates gave in general larger posterior probabilities for the assigned cluster labels compared to clustering without covariates, indicating a superior performance.
Still it is an open question how to optimally weight the importance of the covariates versus the functional data. One might use prior expertise knowledge when available.
It would also be interesting to investigate if the average entropy over the posterior probability distributions could be used for this purpose.

For the functional data, it is assumed that the variance of the measurement error is constant and independent over time. A natural extension would be to allow for a correlation structure for the measurement error within subjects.
Warping is important for aligning the curves. Would it be possible to incorporate warping in the modelling structure?



\clearpage
\appendix
\section{Derivation of parameter estimates for the EM algorithm}
In this Appendix the needed expressions for estimating the parameters within the EM algorithm are derived. First all expressions for functional data without covariates and then in Appendix A.2 the covariates are included to be estimated in the EM algorithm.
\subsection{Functional model without covariates}
\begin{flushleft}
The expected value of the complete log likelihood $\sum_{i=1}^N \log f\left(\boldsymbol{y}_i,\boldsymbol{\gamma}_i,\boldsymbol{z}_i \right)$ in equation (\ref{eqn3})
given the observed data $\boldsymbol{Y} = (\boldsymbol{y}_1, ...., \boldsymbol{y}_N)$ and starting parameter values
$\theta^{[0]} =( \boldsymbol{\pi}^{[0]},  \Gamma_1^{[0]}, ...,\Gamma_G^{[0]},\sigma^2_{[0]},\boldsymbol{\lambda}_0^{[0]}, \Lambda^{[0]}, {\boldsymbol{\alpha}}_1^{[0]}, ...,{\boldsymbol{\alpha}}_h^{[0]})$ will first be computed,
noting that, due to independence
\begin{align}
\pi_{k|i}^{[0]} = E [z_{ik}|\boldsymbol{y}_i,\theta^{[0]}] = P(z_{ik}=1|\boldsymbol{y}_i, \theta^{[0]}) = \frac{f(\boldsymbol{y}_i|z_{ik}=1) \pi_k^{[0]}}{\sum_{j=1}^{G}f(\boldsymbol{y}_i|z_{ij}=1)\pi_j^{[0]}}\label{piikgiveni},
\end{align}
where by (\ref{eqn1})
\begin{align}
f(\boldsymbol{y}_i|z_{ik}=1, \theta^{[0]})\sim N_{n_i}(\phi_i (\boldsymbol{\lambda}_0^{[0]}+\Lambda^{[0]}\boldsymbol{\alpha}_k^{[0]}), \sigma_{[0]}^2\mathbb{I}_{n_i}+\phi_i\Gamma_k^{[0]} \phi_i^T). \label{ycond1}
\end{align}

Furthermore, for any function $g(\cdot)$,
\begin{align}
E[z_{ik}g(\boldsymbol{\gamma}_i)|\boldsymbol{y}_i, \theta^{[0]}] = E[g(\boldsymbol{\gamma}_i)|\boldsymbol{y}_i,z_{ik}=1, \theta^{[0]}] \pi_{k|i}^{[0]}.
\end{align}
Hence, the expected value of (\ref{eqn3}) given $\boldsymbol{Y}$ and $\theta^{[0]}$ is, due to independence between individuals
\begin{align}
E[l(\cdot)|\boldsymbol{Y}, \theta^{[0]}] & = \sum_{i=1}^{N}\sum_{k=1}^{G}\pi_{k|i}^{[0]}\log(\pi_{k}) \nonumber\\
& -\frac{1}{2}\sum_{i=1}^{N}\sum_{k=1}^{G}\pi_{k|i}^{[0]}\left[  \log\left\vert \Gamma_{k}\right\vert
+E[\boldsymbol{\gamma}_{i}^{T}\Gamma_{k}^{-1}\boldsymbol{\gamma}_{i}|\boldsymbol{y}_i, z_{ik}=1, \theta^{[0]}]\right]\nonumber\\
& -\frac{1}{2}\sum_{i=1}^{N}\sum_{k=1}^{G}\pi_{k|i}^{[0]}\nonumber\\
&\left[  n_{i}\log\sigma
^{2}+\frac{1}{\sigma^{2}}E\left[\left\vert \left\vert \boldsymbol{y}_{i}-\phi_{i}%
(\boldsymbol{\lambda}_{0}+\Lambda\boldsymbol{\alpha}_{k}+\boldsymbol{\gamma}_{i})\right\vert \right\vert ^{2} | \boldsymbol{y}_i, z_{ik}=1, \theta^{[0]}\right]\right] \label{Ethe rest}
\end{align}
The next step in the EM algorithm to maximize the conditional expectation with respect to the parameters.
Maximizing (\ref{Ethe rest}) with respect to $\pi_1,...,\pi_G$ given that $\sum_{k=1}^G \pi_k = 1$ is equivalent to maximizing
\[
Q_1 = \sum_{i=1}^N \sum_{k=1}^G \pi_{k|i}^{[0]} \log(\pi_k) - {\kappa}(\sum_{k=1}^{G}\pi_k - 1).
\]
Taking derivatives yields
\[
\frac{dQ_1}{d\pi_k} = \frac{1}{\pi_k}\sum_{i=1}^{N}\pi_{k|i}^{[0]}-{\kappa}=0,
 \hspace{3mm} k=1,...,G,
\]
and thus
\[
\hat{\pi}_k^{[1]}= \frac{1}{\kappa}\sum_{i=1}^{N}\pi_{k|i}^{[0]}, \hspace{3mm} k=1,...,G.
\]
Knowing that $\sum_{k=1}^G \hat{\pi}_k^{[1]} = 1$ we have that $\kappa = \sum_{k=1}^G \sum_{i=1}^N \pi_{k|i}^{[0]} = N$ and thus that the updated parameter estimates are
\begin{align}
\hat{\pi}_{k}^{[1]} = \frac{1}{N}\sum_{i=1}^{N} \pi_{k|i}^{[0]} \label{Piina}
,\hspace{3mm}k=1,...,G.
\end{align}
\vspace{5mm}
In order to maximize (\ref{Ethe rest}) with respect to $\Gamma_1, ..., \Gamma_K$, first note that
\begin{align}
E [\boldsymbol{\gamma}_{i}^{T}\Gamma_{k}^{-1}\boldsymbol{\gamma}_{i}] = E [tr(\boldsymbol{\gamma}_{i}^{T}\Gamma_{k}^{-1}\boldsymbol{\gamma}_{i}) ] = E [tr(\Gamma_{k}^{-1}\boldsymbol{\gamma}_{i}\boldsymbol{\gamma}_{i}^{T})] = tr(\Gamma_{k}^{-1} E[\boldsymbol{\gamma}_{i}\boldsymbol{\gamma}_{i}^{T}])\label{trace}
\end{align}
always holds.
Since there is a one to one relation between $\Gamma_k$ and $\Gamma_k^{-1}$ we maximize (\ref{Ethe rest}) with respect to $\Gamma_k^{-1}$ which by (\ref{trace}) is equivalent to maximizing
\begin{align}
Q_2 = -\frac{1}{2}\sum_{i=1}^{N}\sum_{k=1}^{G}\pi_{k|i}^{[0]}\left[ - \log\left\vert \Gamma_{k}^{-1}\right\vert
+ tr(\Gamma_{k}^{-1} E[\boldsymbol{\gamma}_{i}\boldsymbol{\gamma}_{i}^{T}|\boldsymbol{y}_i, z_{ik}=1,\theta^{[0]}])\right],\label{E0}
\end{align}
with respect to $\Gamma_k^{-1}$ where we have used the fact that $\log\left\vert \Gamma_{k}\right\vert = - \log\left\vert \Gamma_{k}^{-1}\right\vert$.
We further have that
\[
\frac{dQ_2}{d\Gamma_k^{-1}} = -\frac{1}{2}\sum_{i=1}^{N}\pi_{k|i}^{[0]} \left( -\Gamma_k + E\left[\boldsymbol{\gamma}_i \boldsymbol{\gamma}_i^T| \boldsymbol{y}_i, z_{ik}=1,\theta^{[0]} \right]\right), \hspace{3mm}k=1,...,G.
\]
Setting the derivative equal to zero and solving the equations gives by (\ref{Piina}) that (\ref{E0}) is maximized by
\begin{align}
\hat{\Gamma}_k^{[1]} 
= \frac{1}{N\hat{\pi}_k^{[0]}}\sum_{i=1}^{N}\pi_{k|i}^{[0]} E\left[\boldsymbol{\gamma}_i \boldsymbol{\gamma}_i^T| \boldsymbol{y}_i, z_{ik}=1, \theta^{[0]}\right], \hspace{3mm}k=1,...,G.\label{GammaK}
\end{align}
We can further express
\begin{align}
E\left[\boldsymbol{\gamma}_i \boldsymbol{\gamma}_i^T| \boldsymbol{y}_i, z_{ik}=1, \theta^{[0]} \right] =
V\left[\boldsymbol{\gamma}_i| \boldsymbol{y}_i, z_{ik}=1,\theta^{[0]} \right] + \hat{\boldsymbol{\gamma}}_{ik}^{[0]} \hat{\boldsymbol{\gamma}}_{ik}^{[0]T}, \label{omskr1}
\end{align}
where
\begin{align}
\hat{\boldsymbol{\gamma}}_{ik}^{[0]} = E\left[\boldsymbol{\gamma}_i|\boldsymbol{y}_i, z_{ik}=1, \theta^{[0]}\right].\label{omskr2}
\end{align}

\end{flushleft}%


\begin{flushleft}
Now, $\boldsymbol{\gamma}_i$ given $\boldsymbol{y}_i, z_{ik}=1$ and $\theta$ follows the $p$-dimensional multivariate normal distribution
\end{flushleft}

\begin{align}
\boldsymbol{\gamma}_{i}|\boldsymbol{y}_{i},z_{ik}=1  &  \sim\nonumber\\
&  N_p((\sigma^{2}\Gamma_{k}^{-1}+\phi_{i}^{T}\phi_{i})^{-1}\phi_{i}
^{T}(\boldsymbol{y} _{i}-\phi_{i}\boldsymbol{\lambda}_{0}-\phi_{i}\Lambda \boldsymbol{\alpha}_k
),(\Gamma_{k}^{-1}+\phi_{i}^{T} \phi_{i}/\sigma^{2})^{-1}),\label{gammagivenyz}
\end{align}
(see e.g Appendix $A.2$, equations (\ref{cond.norm})--(\ref{cond.exp}) for a similar justification).
\begin{flushleft}
Hence, $\hat{\Gamma}_k^{[1]}$ in (\ref{GammaK}) can be calculated using (\ref{omskr2}) and (\ref{omskr1}) with the mean and variance given by (\ref{gammagivenyz}) with $\theta = \theta^{[0]}$.
\end{flushleft}
 Next (\ref{Ethe rest}) is maximized  with respect to $\boldsymbol{\lambda}_0, \Lambda, \boldsymbol{\alpha}_k, k=1,...,G$ which is equivalent to maximizing
\[
Q_3 = -\frac{1}{2}\sum_{i=1}^{N}\sum_{k=1}^{G}\pi_{k|i}^{[0]}\left[  n_{i}\log\sigma
^{2}+\frac{1}{\sigma^{2}}E\left[\left\vert \left\vert \boldsymbol{y}_{i}-\phi_{i}%
(\boldsymbol{\lambda}_{0}+\Lambda\boldsymbol{\alpha}_{k}+\boldsymbol{\gamma}_{i})\right\vert \right\vert ^{2} | \boldsymbol{y}_i, z_{ik}=1,\theta^{[0]}\right]\right].
\]
\begin{flushleft}
As in \cite{james}, the maximization of $Q_3$ will be an iterative procedure where $Q_3$ is first maximized with respect to $\boldsymbol{\lambda}_0$ then to $\boldsymbol{\alpha}_k, k=1,..,G$, and finally to the columns of $\Lambda$ repeatedly while holding all other parameters fixed.
We start by maximizing $Q_3$ with respect to $\boldsymbol{\lambda}_0$, which implies maximizing
\begin{align}
Q_3(\boldsymbol{\lambda}_0) &= -\sum_{i=1}^{N}\sum_{k=1}^{G}\pi_{k|i}^{[0]}
 E\left[(c_{ik}-\phi_i \boldsymbol{\lambda}_0)^T (c_{ik}-\phi_i \boldsymbol{\lambda}_0) | \boldsymbol{y}_i, z_{ik}=1, \theta^{[0]}\right]\\
&= -\sum_{i=1}^{N}\sum_{k=1}^{G}\pi_{k|i}^{[0]} \left[ E[c_{ik}^Tc_{ik}| \boldsymbol{y}_i, z_{ik}=1, \theta^{[0]}] - 2 E[c_{ik}^T| \boldsymbol{y}_i, z_{ik}=1, \theta^{[0]}]\phi_i \boldsymbol{\lambda}_0
+ \boldsymbol{\lambda}_0^T \phi_i^T \phi_i \boldsymbol{\lambda}_0 \right],
\end{align}
where $c_{ik} = \boldsymbol{y}_i - \phi_i \Lambda \boldsymbol{\alpha}_k - \phi_i \boldsymbol{\gamma}_i$.
Taking the derivative of $Q_3$ with respect to $\boldsymbol{\lambda}_0$ we get
\begin{align*}
\frac{dQ_3(\boldsymbol{\lambda}_0)}{d\boldsymbol{\lambda}_0} =
-\sum_{i=1}^{N}\sum_{k=1}^{G}\pi_{k|i}^{[0]} \left[(-2)\phi_i^T E[c_{ik} | \boldsymbol{y}_i, z_{ik}=1, \theta^{[0]}] +  2 \phi_i^T \phi_i \boldsymbol{\lambda}_0 \right],
\end{align*}
which by setting it equal to zero, by (\ref{omskr2}) gives the updated parameter estimate
\begin{align}
 \Rightarrow {\boldsymbol{\lambda}}_0^{[1]} &= \left( \sum_{i=1}^{N} \phi_i^T \phi_i \right)^{-1} \sum_{i=1}^{N}\sum_{k=1}^{G}\pi_{k|i}^{[0]} \phi_i^T E[c_{ik} | \boldsymbol{y}_i, z_{ik}=1, \theta^{[0]}]\nonumber\\
&= \left( \sum_{i=1}^{N} \phi_i^T \phi_i \right)^{-1}
\sum_{i=1}^{N} \phi_i^T \left[ \boldsymbol{y}_i - \sum_{k=1}^{G}\pi_{k|i}^{[0]} \phi_i (\Lambda_0 \boldsymbol{\alpha}_k + \hat{\boldsymbol{\gamma}}_{ik}^{[0]})\right]. \label{lambdazero}
\end{align}
where $\Lambda$ and $\boldsymbol{\alpha}_k$, $k=1,...,G$ in (\ref{lambdazero}) are replaced by $\Lambda^{[0]}$ and $\boldsymbol{\alpha}^{[0]}_k$, $k=1,...,G$ respectively.

To maximize $Q_3$ with respect to $\boldsymbol{\alpha}_k$ it is sufficient to maximize
\begin{align*}
Q_3(\boldsymbol{\alpha}_k) &= -\sum_{i=1}^{N} \pi_{k|i}^{[0]} E[(g_i - \phi_i \Lambda \boldsymbol{\alpha}_k)^T(g_i - \phi_i \Lambda \boldsymbol{\alpha}_k)| \boldsymbol{y}_i, z_{ik}=1, \theta^{[0]} ]\\
&= -\sum_{i=1}^{N} \pi_{k|i}^{[0]} \left[ E[g_i^T g_i| \boldsymbol{y}_i, z_{ik}=1, \theta^{[0]}] - 2 E[g_i^T| \boldsymbol{y}_i, z_{ik}=1, \theta^{[0]}]\phi_i \Lambda \boldsymbol{\alpha}_k + \boldsymbol{\alpha}_k^T \Lambda^T \phi_i^T \phi_i \Lambda \boldsymbol{\alpha}_k \right],
\end{align*}
where $g_i=\boldsymbol{y}_i - \phi_i (\boldsymbol{\lambda}_0+\boldsymbol{\gamma}_i)$.
The derivative of $Q_3(\boldsymbol{\alpha}_k)$ with respect to $\boldsymbol{\alpha}_k$ is
\begin{align*}
\frac{dQ_3(\boldsymbol{\alpha}_k)}{d\boldsymbol{\alpha}_k} &= -\sum_{i=1}^{N} \pi_{k|i}^{[0]} (-2 \Lambda^T \phi_i^T E[g_i| \boldsymbol{y}_i, z_{ik}=1, \theta^{[0]}] + 2 \Lambda^T \phi_i^T \phi_i \Lambda \boldsymbol{\alpha}_k), \nonumber\\
\end{align*}
which beeing set equal to zero by (\ref{omskr2}) yields the updated parameter estimate
\begin{align}
{\boldsymbol{\alpha}}_k^{[1]} &= \left(\sum_{i=1}^{N} \pi_{k|i}^{[0]} \Lambda^{[0]T} \phi_i^T \phi_i \Lambda^{[0]} \right)^{-1} \sum_{i=1}^{N} \pi_{k|i}^{[0]} \Lambda^{[0]T} \phi_i^T ( \boldsymbol{y}_i - \phi_i \boldsymbol{\lambda}_0^{[0]} - \phi_i \hat{\boldsymbol{\gamma}}_{ik}^{[0]} ), k=1,...,G. \label{alphak}
\end{align}

We now maximize $Q_3$ for each column of $\Lambda = (\boldsymbol{\lambda}_1, \boldsymbol{\lambda}_2, ...,\boldsymbol{\lambda}_h)$. To maximize $Q_3$ with respect to $\boldsymbol{\lambda}_m$ it is sufficient to maximize
{\small
\begin{align*}
Q_3(\boldsymbol{\lambda}_m) &= -\sum_{i=1}^{N}\sum_{k=1}^{G}\pi_{k|i}^{[0]} E[(\boldsymbol{e}_{ikm}-\phi_i \alpha_{km} \boldsymbol{\lambda}_m)^T(\boldsymbol{e}_{ikm}-\phi_i {\alpha}_{km} \boldsymbol{\lambda}_m)| \boldsymbol{y}_i, z_{ik}=1, \theta^{[0]}]\\
&= -\sum_{i=1}^{N}\sum_{k=1}^{G}\pi_{k|i}^{[0]} \left[ E[\boldsymbol{e}_{ikm}^T \boldsymbol{e}_{ikm} | \boldsymbol{y}_i, z_{ik}=1, \theta^{[0]}] -2 E[\boldsymbol{e}_{ikm}^T| \boldsymbol{y}_i, z_{ik}=1, \theta^{[0]}] \phi_i {\alpha}_{km} \boldsymbol{\lambda}_m
  +\boldsymbol{\lambda}^T_m {\alpha}_{km}^2 \phi_i^T \phi_i \boldsymbol{\lambda}_m\right]
\end{align*}
}
where $\boldsymbol{e}_{ikm} = \boldsymbol{y}_i - \phi_i (\boldsymbol{\lambda}_0 -\boldsymbol{\gamma}_i -\sum_{l \neq m} \boldsymbol{\lambda}_l {\alpha}_{kl})$ and
$\boldsymbol{\alpha}_k= (\alpha_{k1},...,\alpha_{kh})^T$.
Next, taking the derivative of $Q_3(\boldsymbol{\lambda}_m)$ with respect to $\boldsymbol{\lambda}_m$ yields
\begin{align*}
\frac{dQ_3(\boldsymbol{\lambda}_m)}{d\boldsymbol{\lambda}_m} &= -\sum_{i=1}^{N} \sum_{k=1}^G \pi_{k|i}^{[0]} (-2 {\alpha}_{km} \phi_i^T E[\boldsymbol{e}_{ikm}| \boldsymbol{y}_i, z_{ik}=1, \theta^{[0]}] + 2 {\alpha}_{km}^2 \phi^T_i \phi_i \boldsymbol{\lambda}_m),
\end{align*}
which, equals to zero, by (\ref{omskr2}) gives
\begin{align}
{\boldsymbol{\lambda}}_m^{[1]} &= \left(\sum_{i=1}^{N} \sum_{k=1}^G \pi_{k|i}^{[0]} ({\alpha}_{km}^{[0]})^2 \phi_i^T \phi_i \right)^{-1} \sum_{i=1}^{N} \sum_{k=1}^{G} \pi_{k|i}^{[0]} {\alpha}_{km}^{[0]} \phi_i^T \left( \boldsymbol{y}_i - \phi_i (\boldsymbol{\lambda}_0^{[0]} -  \sum_{l\neq m}{\alpha}_{kl}^{[0]} \boldsymbol{\lambda}_l^{[0]} -\hat{\boldsymbol{\gamma}}_{ik}^{[0]} )\right). \label{Lambdam}
\end{align}

Finally maximize $Q_3$ with respect to $\sigma^2$ is equivalent to maximizing
\begin{align*}
Q_3(\sigma^2) &= -\sum_{i=1}^{N}\sum_{k=1}^{G}\pi_{k|i}^{[0]} \left[n_i \log(\sigma^2)+\frac{1}{\sigma^2} \tau_{ik}
\right]
\end{align*}
where $\tau_{ik}= E\left[ \left\vert \left\vert \boldsymbol{y}_{i}-\phi_{i}%
(\boldsymbol{\lambda}_{0}+\Lambda\boldsymbol{\alpha}_{k}+\boldsymbol{\gamma}_{i}) \right\vert \right\vert ^{2} | \boldsymbol{y}_i, z_{ik}=1, \theta^{[0]} \right]$.
The derivative with respect to $\sigma^2$ is
\begin{align*}
\frac{dQ_3(\sigma^2)}{d\sigma^2} &= \sum_{i=1}^{N}\sum_{k=1}^{G}\pi_{k|i}^{[0]} \left[\frac{n_i}{\sigma^2} -\frac{1}{(\sigma^2)^2} \tau_{ik}
\right]
\end{align*}
which equal to zero, yields
\begin{align*}
{\sigma}^2 = \frac{1}{\sum_{i=1}^{N}n_i} \sum_{i=1}^{N}\sum_{k=1}^{G}\pi_{k|i}^{[0]} \tau_{ik}
\end{align*}
Note that $\tau_{ik}=\tau_{ik}(\boldsymbol{\lambda}_0,\Lambda,\boldsymbol{\alpha}_k)$ can be computed as follows. By letting
\begin{align*}
r_{ik} = \boldsymbol{y}_i - \phi_i \boldsymbol{\lambda}_0 - \phi_i \Lambda \boldsymbol{\alpha}_k
\end{align*}
we have that 
\begin{align}
\tau_{ik} 
= r_{ik}^T r_{ik} -2 r_{ik}^T \phi_i \hat{\boldsymbol{\gamma}}_{ik}^{[0]} + E\left[\boldsymbol{\gamma}_i^T\phi_i^T \phi_i  \boldsymbol{\gamma}_i| \boldsymbol{y}_i, z_{ik}=1, \theta^{[0]} \right]
\label{tauik}.
\end{align}
By (\ref{alphak}) and (\ref{Lambdam}) it further holds that
\begin{align*}
E\left[\boldsymbol{\gamma}_i^T\phi_i^T \phi_i  \boldsymbol{\gamma}_i| \boldsymbol{y}_i, z_{ik}=1, \theta^{[0]} \right] &
&= tr(\phi_i V\left[\boldsymbol{\gamma}_i|\boldsymbol{y}_i, z_{ik}=1, \theta^{[0]}\right]\phi_i^T) + tr(\phi_i \hat{\gamma}_{ik}^{[0]} \hat{\gamma}_{ik}^{[0]T}\phi_i^T).
\end{align*}
Note that
\begin{align*}
V[\boldsymbol{\gamma}_i | \boldsymbol{y}_i, z_{ik}=1, \theta^{[0]} ] &= E[(\boldsymbol{\gamma}_i - \hat{\boldsymbol{\gamma}}_{ik})^T(\boldsymbol{\gamma}_i - \hat{\boldsymbol{\gamma}}_{ik})| \boldsymbol{y}_i, z_{ik}=1, \theta^{[0]} ]\\
&= E[\boldsymbol{\gamma}_i \boldsymbol{\gamma}_i^T| \boldsymbol{y}_i, z_{ik}=1, \theta^{[0]} ] -\hat{\boldsymbol{\gamma}}_{ik}\hat{\boldsymbol{\gamma}}_{ik}^T.
\end{align*}
It then gives us by (\ref{tauik}) that
\begin{align*}
\tau_{ik}(\boldsymbol{\lambda}_0,\Lambda,\boldsymbol{\alpha}_k)
&= (r_{ik}-\phi_i \hat{\boldsymbol{\gamma}}_{ik}^{[0]})^T (r_{ik}-\phi_i \hat{\boldsymbol{\gamma}}_{ik}^{[0]}) + tr(\phi_i
V\left[\boldsymbol{\gamma}_i| \boldsymbol{y}_i, z_{ik}=1, \theta^{[0]} \right] \phi_i^T)
\end{align*}
where $ V\left[\boldsymbol{\gamma}_i| \boldsymbol{y}_i, z_{ik}=1, \theta^{[0]} \right]$ is the covariance matrix of (\ref{gammagivenyz}).
The updated parameter estimate of $\sigma^2$ then becomes
\begin{align}
{\hat{\sigma}^2_{[1]}} = \frac{1}{\sum_{i=1}^{N}n_i} \sum_{i=1}^{N}\sum_{k=1}^{G}\hat{\pi}_{k|i}^{[0]} \tau_{ik}(\boldsymbol{\lambda}_0^{[0]},\Lambda^{[0]},\boldsymbol{\alpha}_k^{[0]}) \label{sigma}
\end{align}
\end{flushleft}
\subsection{Functional model with covariates}
\begin{flushleft}
Covariates $\boldsymbol{x}_i$, can also be included together with the functional data when clustering individuals. The expected value of the complete log likelihood (6) including covariates given $\boldsymbol{U}$ and $\theta^{[0]}$ then becomes
\begin{align}
E [l(\theta)|\boldsymbol{U}, \theta^{[0]}] & \propto \sum_{i=1}^{N}\sum_{k=1}^{G}\tilde{\pi}_{k|i}^{[0]}\log(\pi_{k})\nonumber\\
& -\frac{1}{2}\sum_{i=1}^{N}\sum_{k=1}^{G}\tilde{\pi}_{k|i}^{[0]}\left[  \log\left\vert \Delta_{k}\right\vert
+E [\boldsymbol{\xi}_{i}^{T}\Delta_{k}^{-1}\boldsymbol{\xi}_{i}|\boldsymbol{u}_i,z_{ik}=1, \theta^{[0]} ]\right]\nonumber\\
& -\frac{1}{2}\sum_{i=1}^{N}\sum_{k=1}^{G}\tilde{\pi}_{k|i}^{[0]}\left[  n_{i}\log\sigma
^{2}+\frac{1}{\sigma^{2}}E\left[\left\vert \left\vert \boldsymbol{y}_{i}-\phi_{i}%
(\boldsymbol{\lambda}_{0}+\Lambda\boldsymbol{\alpha}_{k}+\boldsymbol{\gamma}_{i})\right\vert \right\vert ^{2} | \boldsymbol{u}_i, z_{ik}=1,\theta^{[0]}\right]\right] \nonumber\\
& -\frac{1}{2}\sum_{i=1}^{N}\sum_{k=1}^{G}\tilde{\pi}_{k|i}^{[0]}\left[ r \log(\sigma^2_x) + \frac{1}{\sigma^{2}_{x}} E\left[ \left\vert \left\vert \boldsymbol{x}_{i} -
(\boldsymbol{\upsilon}_{j}+\boldsymbol{\delta}_{i})\right\vert \right\vert^2 | \boldsymbol{u}_i, z_{ik}=1, \theta^{[0]}\right]\right], \label{app.covar}
\end{align}
where
\[
\tilde{\pi}_{k|i}^{[0]} = E [z_{ik}|\boldsymbol{u}_i,\theta^{[0]}] 
= \frac{f_0(\boldsymbol{u}_i|z_{ik}=1) \pi_k^{[0]}}{\sum_{j=1}^{G}f_0(\boldsymbol{u}_i|z_{ij}=1)\pi_j^{[0]}},
\]
and
\begin{align}
f_0(\boldsymbol{u}_i|z_{ik}=1)\sim N_{n_i+r}(S_i \boldsymbol{\mu}_k^{[0]}, R^{[0]}_{(n_i+r)}+S_i \Delta_k^{[0]} S_i^{T}). \label{app.ycond1}
\end{align}

Maximizing (\ref{app.covar}) with respect to $\pi_k$, $k=1,...,G$ follows the same argument as when estimating the $\pi_k$:s without covariates as in Appendix A.1 pages 23-24, changing the conditioning from $\boldsymbol{y}_i$ to $\boldsymbol{u}_i$ giving the updated parameter estimates
\[
\hat{\pi}_{k}^{[1]} = \frac{1}{N}\sum_{i=1}^{N} \tilde{\pi}_{k|i}^{[0]}, \hspace{3mm} k=1,...,G.
\]
Moreover, the $\Delta_k$ that maximizes (\ref{app.covar}) satisfies
\begin{align*}
\hat{\Delta}_k^{[1]} = \frac{1}{\sum_{i=1}^N \tilde{\pi}_{k|i}^{[0]}}\sum_{i=1}^{N}\tilde{\pi}_{k|i}^{[0]} E\left[\boldsymbol{\xi}_i \boldsymbol{\xi}_i^T |\boldsymbol{u}_i, z_{ik}=1,\theta{[0]}\right]
\hspace{3mm}k=1,...,G,
\end{align*}
where
\begin{align*}
E\left[\boldsymbol{\xi}_i \boldsymbol{\xi}_i^T |\boldsymbol{u}_i, z_{ik}=1,\theta{[0]}\right] =
V\left[\boldsymbol{\xi}_i|\boldsymbol{u}_i,z_{ik}=1, \theta^{[0]} \right] + \hat{\boldsymbol{\xi}}_{ik} \hat{\boldsymbol{\xi}}_{ik}^T
\end{align*}
and
\begin{align*}
\hat{\boldsymbol{\xi}}_{ik} = E\left[\boldsymbol{\xi}_i|\boldsymbol{u}_i,z_{ik}=1, \theta^{[0]} \right],
\end{align*}
motivated in a similiar way as equations (\ref{lambdazero})-(\ref{Lambdam}).
\end{flushleft}

\begin{flushleft}
Note that, in order to fully compute $\hat{\Delta}_k^{[1]}$ we need to determine the conditional distribution of $\boldsymbol{\xi}_i$ given $\boldsymbol{u}_i$ and $z_{ik}=1$. First, since both $\boldsymbol{\xi}_i$ and $\boldsymbol{u}_i$ are normally distributed so is $\boldsymbol{\xi}_i$ given $\boldsymbol{u}_i$ with density
\begin{align}
f(\boldsymbol{\xi}_{i} | \boldsymbol{u}_{i}) 
\propto f(\boldsymbol{\xi}_{i}) f(\boldsymbol{u}_{i}|\boldsymbol{\xi}_{i})\label{cond.norm}
\end{align}

For any multivariate normally distributed random variable $\boldsymbol{x}$ with mean $\boldsymbol{a}$ and variance-covariance matrix $V$, the density function equals
\begin{align}
f(\boldsymbol{x}) = C \exp(-\frac{1}{2}(\boldsymbol{x}-\boldsymbol{a})^T V^{-1}(\boldsymbol{x}-\boldsymbol{a})) \propto
\exp\left[-\frac{1}{2}\left(\boldsymbol{x}^T V^{-1} \boldsymbol{x} - 2 \boldsymbol{a}^T V^{-1} \boldsymbol{x} + \boldsymbol{a}^T V^{-1}\boldsymbol{a}\right)\right] \label{mult.norm.exp},
\end{align}
where $C$ is a normalizing constant.
Given $z_{ik}=1$ we have that  $\boldsymbol{\xi}_i \sim N(\boldsymbol{0},\Delta_k)$ and $\boldsymbol{u}_i|\boldsymbol{\xi}_i \sim N(S_i (\boldsymbol{\mu}_k+\boldsymbol{\xi}_i),R)$. By combining (\ref{cond.norm}), (\ref{mult.norm.exp}) we have that
\begin{align}
-2 \ln f(\boldsymbol{\xi}_i|\boldsymbol{u}_i) = C_1 +
\boldsymbol{\xi}_i^T \Delta_k^{-1}\boldsymbol{\xi}_i + (\boldsymbol{A} - S_i \boldsymbol{\xi}_i)^T R^{-1} (\boldsymbol{A} - S_i \boldsymbol{\xi}_i), \label{new.expr}
\end{align}
where $C_1$ is some constant and $\boldsymbol{A}=\boldsymbol{u}_i - S_i \boldsymbol{\mu}_{k}$. Rearranging terms in (\ref{new.expr}) yields
\begin{eqnarray*}
\boldsymbol{\xi}_i^T (\Delta_k^{-1} + S_i^{T} R^{-1} S_i) \boldsymbol{\xi}_i + \boldsymbol{A}^T R^{-1} \boldsymbol{A} - 2(\boldsymbol{A}^T R^{-1} S_i)
\boldsymbol{\xi}_i,
\end{eqnarray*}
which, by comparison with(\ref{mult.norm.exp}) yields
\begin{align}
Var\left[\boldsymbol{\xi}_i| \boldsymbol{u}_i,z_{ik}=1 \right] = V = (\Delta_k^{-1} + S_i^{T} R^{-1} S_i)^{-1}, \label{cond.var}
\end{align}
and
\begin{align}
E\left[\boldsymbol{\xi}_i|\boldsymbol{u}_i,z_{ik}=1 \right] = \left(S_i^T R^{-1} S_i +\Delta_k^{-1} \right)^{-1} S_i^T R^{-1}(\boldsymbol{u}_i-S_i \boldsymbol{\mu}_k).\label{cond.exp}
\end{align}
\end{flushleft}
\begin{flushleft}
To maximize (\ref{app.covar}) with respect to $\boldsymbol{\lambda}_0, \Lambda, \boldsymbol{\alpha}_k$ and $\sigma^2$ is equivalent to maximizing
\[
Q_3 = -\frac{1}{2}\sum_{i=1}^{N}\sum_{k=1}^{G}\tilde{\pi}_{k|i}^{[0]}\left[  n_{i}\log\sigma
^{2}+\frac{1}{\sigma^{2}}E\left[\left\vert \left\vert \boldsymbol{y}_{i}-\phi_{i}%
(\boldsymbol{\lambda}_{0}+\Lambda\boldsymbol{\alpha}_{k}+\boldsymbol{\gamma}_i)\right\vert \right\vert ^{2} | \boldsymbol{u}_i,z_{ik}=1, \theta^{[0]} \right]\right].
\]

Maximizing $Q_3$ will be an iterative maximization where first $\boldsymbol{\lambda}_0$ then $\boldsymbol{\alpha}_k$ and finally the columns of $\Lambda$ are repeatedly maximized while holding all other parameters fixed. 

In analogy with the motivation of (\ref{lambdazero}) we have that
\begin{align*}
 \hat{\boldsymbol{\lambda}}_0^{[1]} &= \left( \sum_{i=1}^{N} \phi_i^T \phi_i \right)^{-1} \sum_{i=1}^{N}\sum_{k=1}^{G}\tilde{\pi}_{k|i}^{[0]} \phi_i^T E[c_{ik} | \boldsymbol{u}_i,z_{ik}=1, \theta^{[0]} ]\\
&= \left( \sum_{i=1}^{N} \phi_i^T \phi_i \right)^{-1}
\sum_{i=1}^{N} \phi_i^T \left[ \boldsymbol{y}_i - \sum_{k=1}^{G}\tilde{\pi}_{k|i}^{[0]} \phi_i (\Lambda^{[0]} \boldsymbol{\alpha}_k^{[0]} + \tilde{\boldsymbol{\gamma}}_{ik}^{[0]})\right],\\
\end{align*}
where $\tilde{\gamma}_{ik}^{[0]}=E\left[\boldsymbol{\gamma}_i|\boldsymbol{u}_i,z_{ik}=1, \theta^{[0]}\right]$ being the first $p$ rows of (\ref{cond.exp}) with $\theta=\theta^{[0]}$.
We further have that $Q_3$ is maximized with respect to $\boldsymbol{\alpha}_k$, $(k=1,...,G)$ for
\begin{align*}
\hat{\boldsymbol{\alpha}}_k^{[1]} &= \left(\sum_{i=1}^{N} \tilde{\pi}_{k|i}^{[0]} \Lambda^{[0]T} \phi_i^T \phi_i \Lambda^{[0]} \right)^{-1} \sum_{i=1}^{N} \tilde{\pi}_{k|i}^{[0]} \Lambda^{[0]T} \phi_i^T ( \boldsymbol{y}_i - \phi_i \boldsymbol{\lambda}^0_0 - \phi_i \tilde{\gamma}_{ik}^{[0]} ),
\end{align*}
with parallell arguments as those for equation(\ref{alphak}).
Maximizing $Q_3$ with respect to each of the columns of $\Lambda = (\boldsymbol{\lambda}_1, \boldsymbol{\lambda}_2, ...,\boldsymbol{\lambda}_h)$ then, as for equation (\ref{Lambdam}) yields
\begin{align*}
\hat{\boldsymbol{\lambda}}_m^{[1]} &= \left(\sum_{i=1}^{N} \tilde{\pi}_{k|i}^{[0]} \boldsymbol{\alpha}_{km}^{2(0)} \phi_i^T \phi_i \right)^{-1} \sum_{i=1}^{N} \sum_{k=1}^{G} \tilde{\pi}_{k|i}^{[0]} \boldsymbol{\alpha}_{km}^0 \phi_i^T \left( \boldsymbol{y}_i - \phi_i (\boldsymbol{\lambda}_0^0 -  \sum_{l\neq m}\boldsymbol{\alpha}_{kl}^0 \boldsymbol{\lambda}_l^0 -\tilde{\gamma}_{ik}^{[0]} )\right), m=1,...,h.
\end{align*}

To maximize $Q_3$ with respect to $\sigma^2$ we maximize
\begin{align*}
Q_3^{(i)}(\sigma^2) &= -\sum_{i=1}^{N}\sum_{k=1}^{G}\tilde{\pi}_{k|i}^{[0]} \left[n_i \log(\sigma^2)+\frac{1}{\sigma^2} \tilde{\tau}_{ik},
\right]
\end{align*}
where $\tilde{\tau}_{ik}= E\left[ \left\vert \left\vert \boldsymbol{y}_{i}-\phi_{i}%
(\boldsymbol{\lambda}_{0}+\Lambda\boldsymbol{\alpha}_{k}+\boldsymbol{\gamma}_{i}) \right\vert \right\vert ^{2} | \boldsymbol{u}_i,z_{ik}=1, \theta^{[0]}  \right]$.
Taking the derivative with respect to $\sigma^2$ yields
\begin{align*}
\frac{dQ_3^{(iv)}}{d\sigma^2} &= \sum_{i=1}^{N}\sum_{k=1}^{G}\tilde{\pi}_{k|i}^{[0]} \left[\frac{n_i}{\sigma^2} -\frac{1}{(\sigma^2)^2} \tilde{\tau}_{ik}
\right] = 0,
\end{align*}
which is equivalent to
\begin{align*}
(\hat{\sigma}^2) = \frac{1}{\sum_{i=1}^{N}n_i} \sum_{i=1}^{N}\sum_{k=1}^{G}\tilde{\pi}_{k|i}^{[0]} \tilde{\tau}_{ik}.
\end{align*}
Similarly to the computation of $\tau_{ik}$ in equation (\ref{sigma}), $\tilde{\tau}_{ik}=\tilde{\tau}_{ik}(\boldsymbol{\lambda}_0,\Lambda,\boldsymbol{\alpha}_k)$ can be computed as
\begin{align*}
\tilde{\tau}_{ik}(\boldsymbol{\lambda}_0,\Lambda,\boldsymbol{\alpha}_k)
&= (r_{ik}-\phi_i \tilde{\gamma}_{ik}^{[0]T})(r_{ik}-\phi_i \tilde{\gamma}_{ik}^{[0]}) +
tr\left(\phi_i V\left[\boldsymbol{\gamma}_i| \boldsymbol{u}_i,z_{ik}=1, \theta^{[0]} \right] \phi_i^T\right),
\end{align*}
where $r_{ik}=(\boldsymbol{y}_i-\phi_i(\boldsymbol{\lambda}_0+\Lambda\boldsymbol{\alpha}_k)$ and
$\tilde{\gamma}_{ik}^{[0]}= E\left[ \boldsymbol{\gamma}_i| \boldsymbol{u}_i, z_{ik}=1, \theta^{[0]}\right]$ is the first $p$ rows of (\ref{cond.exp}) with $\theta=\theta^{[0]}$ and
$V\left[\boldsymbol{\gamma}_i| \boldsymbol{u}_i,z_{ik}=1, \theta^{[0]} \right]$ is the $p \times p$ matrix in the upper left corner of variance-covariance matrix (\ref{cond.var}).
The updated parameter estimate of $\sigma^2$ then becomes
\begin{align*}
{(\hat{\sigma}^2)^{[1]}} = \frac{1}{\sum_{i=1}^{N}n_i} \sum_{i=1}^{N}\sum_{k=1}^{G}\tilde{\pi}_{k|i}^{[0]} \tilde{\tau}_{ik}(\boldsymbol{\lambda}_0^{[0]},\Lambda^{[0]},\boldsymbol{\alpha}_k^{[0]}).
\end{align*}

%
%
Finally we maximize (\ref{app.covar}) with respect to $\boldsymbol{\upsilon}_{k}$ and $\sigma_x^2$, which is equivalent to maximizing
\begin{align*}
 Q_4 = & -\frac{1}{2}\sum_{i=1}^{N}\sum_{k=1}^{G}\tilde{\pi}_{k|i}^{[0]}\left[ r \log(\sigma^2_x) +
 \frac{1}{\sigma^{2}_{x}}  E\left[ \left\vert \left\vert \boldsymbol{x}_{i} -
(\boldsymbol{\upsilon}_{k}+\boldsymbol{\delta}_{i})\right\vert \right\vert^2 | \boldsymbol{u}_i, z_{ik}=1, \theta^{[0]}\right]\right].
\end{align*}
We start by taking the derivative of $Q_4$ with respect to  $\boldsymbol{\upsilon}_{k}$, giving then
\begin{align*}
\frac{dQ_4}{d{\boldsymbol{\upsilon}}_{k}}= \sum_{i=1}^{N}\tilde{\pi}_{k|i}^{[0]}\frac{1}{\sigma_x^2}
\left(2 \boldsymbol{\upsilon}_k -2 (\boldsymbol{x}_i - \tilde{\delta}_{ik}^{[0]})\right),
\end{align*}
where $\tilde{\delta}_{ik}^{[0]} =E\left[\boldsymbol{\delta}_i|\boldsymbol{u}_i, z_{ik}=1,\theta^{[0]} \right]$, corresponds to the last $r$ elements in (\ref{cond.exp}) with $\theta=\theta^{[0]}$.
Setting the derivative equal to zero and solving for $\boldsymbol{\upsilon}_k$ giving the updated parameter estimate of $\boldsymbol{\upsilon}_k$ as
\begin{align*}
\boldsymbol{\upsilon}_{k}^{[1]}= \frac{1}{\sum_{i=1}^{N}\tilde{\pi}_{k|i}^{[0]}}\sum_{i=1}^{N}\tilde{\pi}_{k|i}^{[0]}(\boldsymbol{x}_{i}-\tilde{\delta}_{ik}^{[0]}) =
\frac{1}{N\pi_{k}^{[0]}}\sum_{i=1}^{N}\tilde{\pi}_{k|i}^{[0]}(\boldsymbol{x}_{i}-\tilde{\delta}_{ik}^{[0]})
\end{align*}

If we set $b_{ik}=E\left[ \left\vert \left\vert \boldsymbol{x}_{i} -
(\boldsymbol{\upsilon}_{k}+\boldsymbol{\delta}_{i})\right\vert \right\vert^2 | \boldsymbol{u}_i, z_{ik}=1, \theta^{[0]}\right]$ then the derivative of $Q_4$ with respect to $\sigma_x^2$ becomes
\begin{align*}
\frac{dQ_4}{d\sigma_{x}^2}= -\frac{1}{2}\sum_{i=1}^{N}\sum_{k=1}^{G}\tilde{\pi}_{k|i}^{[0]}\left[\frac{r}{\sigma_x^2}-\frac{1}{(\sigma_x^2)^2} b_{ik}\right]
\end{align*}
which set to zero implies that, where $\sigma^2_x = \sigma^2_x(\boldsymbol{\upsilon})$, satisfies
\begin{align*}
\frac{r}{\sigma_x^2}\sum_{i=1}^{N}\sum_{k=1}^{G}\tilde{\pi}_{k|i}^{[0]} =
\frac{1}{(\sigma_x^2)^2}\sum_{i=1}^{N}\sum_{k=1}^{G}\tilde{\pi}_{k|i}^{[0]} b_{ik}
\end{align*}
giving
\begin{align*}
{\sigma}_x^2 =
\frac{1}{Nr}\sum_{i=1}^{N}\sum_{k=1}^{G}\tilde{\pi}_{k|i}^{[0]} E\left[ \left\vert \left\vert \boldsymbol{x}_{i} -
(\boldsymbol{\upsilon}_{k}+\boldsymbol{\delta}_{i})\right\vert \right\vert^2 | \boldsymbol{u}_i, z_{ik}=1, \theta^{[0]}\right].
\end{align*}
Now, in order to find $\sigma^2_x$ we need to find an estimate of $b_{ik}$. Let $d_{ik} = \boldsymbol{x}_i-\boldsymbol{v}_k$ then we can formulate
\begin{align*}
b_{ik} =
E\left[
\vert \vert
d_{ik} - \boldsymbol{\delta}_i
\vert \vert^2
| \boldsymbol{u}_i, z_{ik}=1, \theta^{[0]}
\right] =
{d}_{ik}^T {d}_{ik} - 2 d_{ik}^T \tilde{\delta}_{ik}^{[0]} +
E\left[\boldsymbol{\delta}_i^T \boldsymbol{\delta}_i| \boldsymbol{u}_i, z_{ik}=1, \theta^{[0]}\right].
\end{align*}
However,
\begin{align*}
E\left[\boldsymbol{\delta}_i^T \boldsymbol{\delta}_i| \boldsymbol{u}_i, z_{ik}=1, \theta^{[0]}\right] =
tr\left(E\left[\boldsymbol{\delta}_i^T \boldsymbol{\delta}_i| \boldsymbol{u}_i, z_{ik}=1, \theta^{[0]}\right]\right) =
tr\left(V\left[\boldsymbol{\delta}_i|\boldsymbol{u}_i, z_{ik}=1, \theta^{[0]}\right]+\tilde{\delta}_{ik}^{[0]}\tilde{\delta}_{ik}^{[0]T}\right)
\end{align*}
Hence
\begin{align*}
b_{ik}(\boldsymbol{\upsilon}_k) =
(d_{ik}- \tilde{\delta}_{ik}^{[0]})^T
(d_{ik}-\tilde{\delta}_{ik}^{[0]})+
tr\left(V\left[\boldsymbol{\delta}_i|\boldsymbol{u}_i, z_{ik}=1, \theta^{[0]}\right]\right),
\end{align*}
where $V\left[\boldsymbol{\delta}_i|\boldsymbol{u}_i, z_{ik}=1, \theta^{[0]}\right]$ equals the $r \times r$ matrix in the lower right corner of (\ref{cond.var}) with $\theta=\theta^{[0]}$. Thus, giving the updated parameter estimates of $\sigma_x^2$ as
\begin{align*}
\left(\hat{\sigma}^2_x\right)^{[1]} = \frac{1}{N} \sum_{i=1}^N \sum_{k=1}^G \tilde{\pi}_{k|i}^{[0]} b_{ik}(\boldsymbol{\upsilon}_k^{[0]})
\end{align*}

\end{flushleft}
\section{EM-algorithm implementation}
\renewcommand{\labelitemi}{$-$}
\renewcommand{\labelitemii}{$-$}
\renewcommand{\labelitemiii}{$-$}
\renewcommand{\labelitemiv}{$-$}
For the data at hand we have run the EM-algorithm in two different settings, without and with three covariates. 


The implementation of the EM-algorithm is done in three functions in R. First an initial step, \textbf{(I)},  which runs once. Then iteration between the E-step and the M-step with the chosen stop value as the criteria for convergence.
The convergence criteria to stop the iterations was set to 0.001 for the absolute relative difference in the estimation of $\sigma^2$ without covariates. With the covariates we used the sum of $\sigma^2$ and $\sigma^2_x$.

The initial step \textbf{(I)} finds the starting values for the EM-algorithm. We started by fitting penalized cubic splines with the $p=8$ B-splines for each seasonal pattern. We penalized the second derivative and used the penalty weight $\boldsymbol{\lambda} = 0.00014625$. Then the initial groupings of the seasonal patterns into $G$ clusters could be done in two ways. One is by running k-means several times on the initial penelized spline coefficients of the data using the pre-specified penalty $\boldsymbol{\lambda}$ and $p$ cubic spline functions with $G$ cluster centroids and keep the best fit.  The second is to uniformly assign a cluster belongings for each subject. Both methods gives the the start values $\hat{\pi}_k$, k=1,...,G. The initial cluster belongings together with the spline coefficients are then used to estimate the initial parameters, $\pi_k^{[0]}$, $k=1,...,G$ as the relative frequencies in the clusters. $\boldsymbol{\lambda}_{0}^{[0]}$ as the average of the spline coefficients, $\boldsymbol{\mu}_k$, $k=1,...,G$ as the average spline coefficients within each cluster. An eigendecomposition of $\boldsymbol{\mu}_1,...,\boldsymbol{\mu}_G$ was the used to initiate $\Lambda^{[0]}$ and ${\boldsymbol{\alpha}_k}^{[0]}$'s. The initial values of the $\Gamma_k$'s where found by setting them all equal to the covariance matrix of the spline coefficients. Finally  ${\sigma^{2}}_{(0)}$ is set equal to $\frac{1}{N}\sum_{i=1}^N||\boldsymbol{y}_i-\phi_i\boldsymbol{\mu}_{z_i}^{[0]}||^2$.

One can note that when the EM-algorithm is implemented some modifications are performed in order to increase the numerical stability. First, according to the model specification we have that
\begin{align*}
\boldsymbol{y}_i = \phi_i \boldsymbol{\eta}_i
\end{align*}
Now, a shift of base is performed, giving instead,
\begin{align*}
\boldsymbol{y}_i = \phi_i \boldsymbol{\eta}_i = U_i D V^T \eta_i = U_i \boldsymbol{\beta}_i
\end{align*}
Where the single value decomposition of $\phi_i$ is used. That gives us
\begin{align*}
\boldsymbol{\beta}_i = D V^T \boldsymbol{\eta}_i
\end{align*}
When we want to transform back to our original $\boldsymbol{\eta}_i$ we have
\begin{align*}
\boldsymbol{\eta}_i = VD^{-1} \boldsymbol{\beta}_i.
\end{align*}
We can see this as a new basis-representation with the difference that all $p$ basis functions are defined over the whole domain, see Figure \ref{basis}. The new basis functions are also orthogonal to each other. Calculating the inverse of expressions involving the $\boldsymbol{U}$-matrix is numerically more stable.

Second, by using the Sherman-Morrison-Woodbury matrix identities the Variance of the coefficients for the splines and the covariates ${\boldsymbol{\xi}_i|\boldsymbol{u}_i,z_{ik}=1}$, (\ref{cond.var}) is estimated as
\begin{align*}
\hat{Var}\left[\boldsymbol{\xi}_i| \boldsymbol{u}_i,z_{ik}=1 \right] = \left(S_i^T R^{-1} S_i +\Delta_k^{-1} \right)^{-1} =\\
\Delta_k - \Delta_k S_i^T (R+ S \Delta_k S_i^T)^{-1} S \Delta_k = \\
\Delta_k - \Delta_k S_i^T R^{-1} (\mathbb{I}+S_i \Delta_k S_i^T R^{-1})^{-1} S_i \Delta_k
\end{align*}
and the same idea goes for estimating the variance for the spline coefficients, $\boldsymbol{\gamma}_i|\boldsymbol{y}_i,z_{ik}=1$, see equation (\ref{gammagivenyz}).

\section{Estimating the distortion function}
In practice, when estimating the distortion function (\ref{dK}), it is suggested in \citet{james} to set $b$ equal to half of the "effective" dimension of the parameters. It is also suggested to replace $\Gamma_{z_i}^{-1}$ with the identity matrix $\mathbb{I}$,
the distortion thus being simplified, according to
\begin{align}
d_G= \frac{1}{p}\min_{c_1, \dots, c_G} E[(\boldsymbol{\eta}_i -\boldsymbol{c}_{z_i})^T (\boldsymbol{\eta}_i -\boldsymbol{c}_{z_i})]\label{dK2}.
\end{align}

\cite{james} also suggest to estimate (\ref{dK2}) by the total within-cluster sum of squares (divided by $N$) of the K-means algorithm applied to the $E[\boldsymbol{\eta}_i|\boldsymbol{y}_i,\hat{\theta}]$'s, where we use an estimate of the expected value of $\boldsymbol{\eta}_i$ given the observed data $\boldsymbol{y}_i$ and the parameter estimate $\hat{\theta}$,
\[
E\left[\boldsymbol{\eta}_i|\boldsymbol{y}_i, \hat{\theta}\right]=E\left[\boldsymbol{\lambda}_i+\Lambda \boldsymbol{\alpha}_{z_i}+\boldsymbol{\gamma}_i|\boldsymbol{y}_i, \hat{\theta}\right]
\]
that is
\[
\hat{\boldsymbol{\eta}}_i = \hat{\boldsymbol{\lambda}}_0+\hat{\Lambda}\sum_{k=1}^G\hat{\boldsymbol{\alpha}}_k \hat{\pi}_{k|i}+\hat{\boldsymbol{\gamma}}_i
\]
where
\[
\hat{\boldsymbol{\gamma}}_i = (\hat{\sigma}^2\sum_{k=1}^G\hat{\Gamma}_k^{-1}\hat{\pi}_{k|i}+\phi_i^T\phi_i)^{-1} \phi_i^T (\boldsymbol{y}_i - \phi_i (\hat{\boldsymbol{\lambda}}_0+\hat{\Lambda}\sum_{k=1}^G\hat{\boldsymbol{\alpha}}_k \hat{\pi}_{k|i}).
\]
Also, in a similar way one can calculate $E[\boldsymbol{\delta}_i|\boldsymbol{u}_i,\hat{\theta}]$.

\begin{figure}[ptb]
\centering
\includegraphics[
height=6in,
width=\textwidth
]{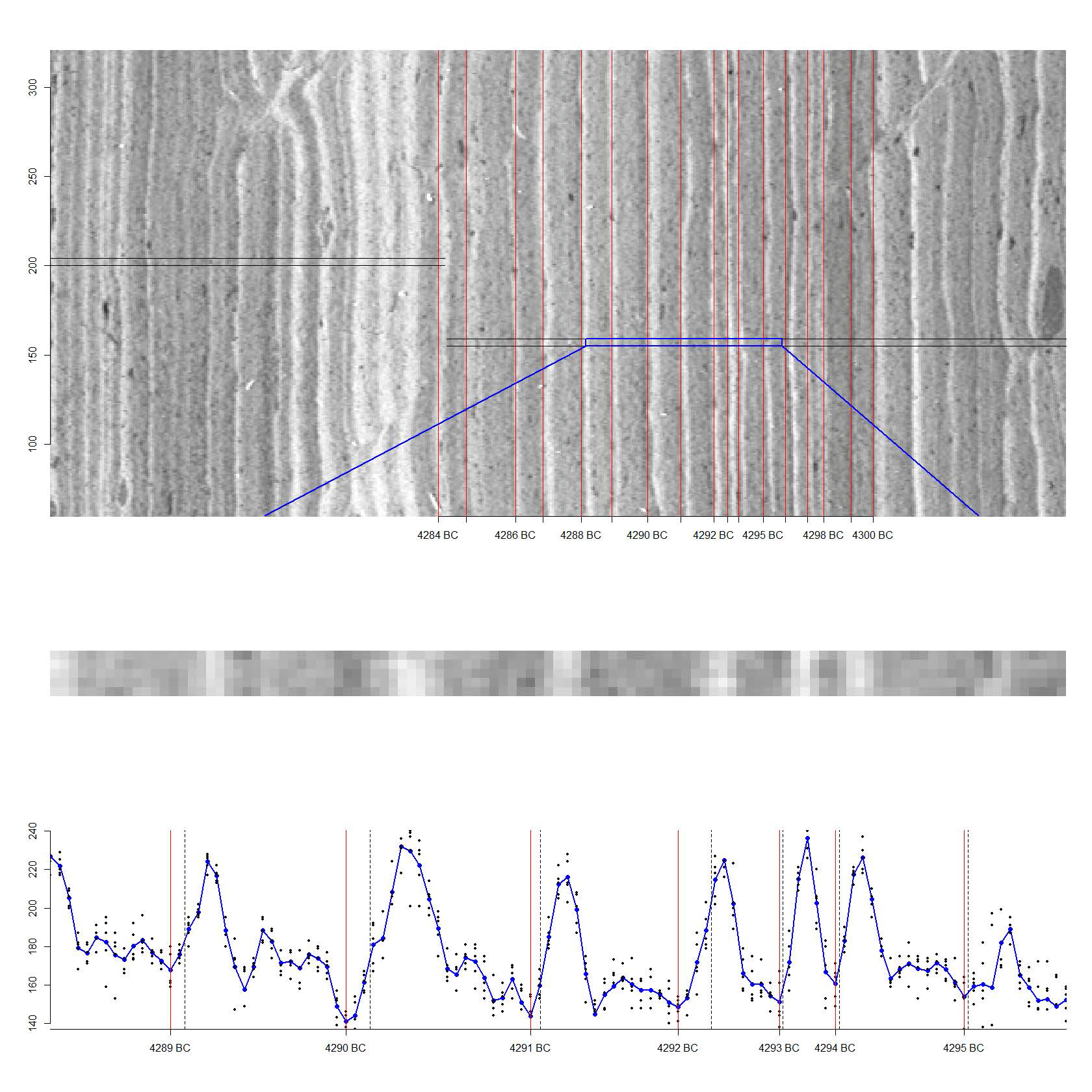}\caption{Annually laminated sediment from lake Kassj\"{o}n (top). Data to be analyzed is based on slices of five pixels width selected from representative parts of the sediment (middle). Gray-scale values for the slice in the middle together with the mean gray-scale values (solid line) of the 5 pixels for each time point (bottom). The manually determined yearly delimiters (black dotted lines) have been horizontally shifted 1-4 steps to the darkest neighboring value (solid red lines).}
\label{illustration}%
\end{figure}
\begin{figure}[ptb]
\centering
\textbf{Determining the number of clusters}\par\medskip
\includegraphics[
height=6in,
width=\textwidth
]{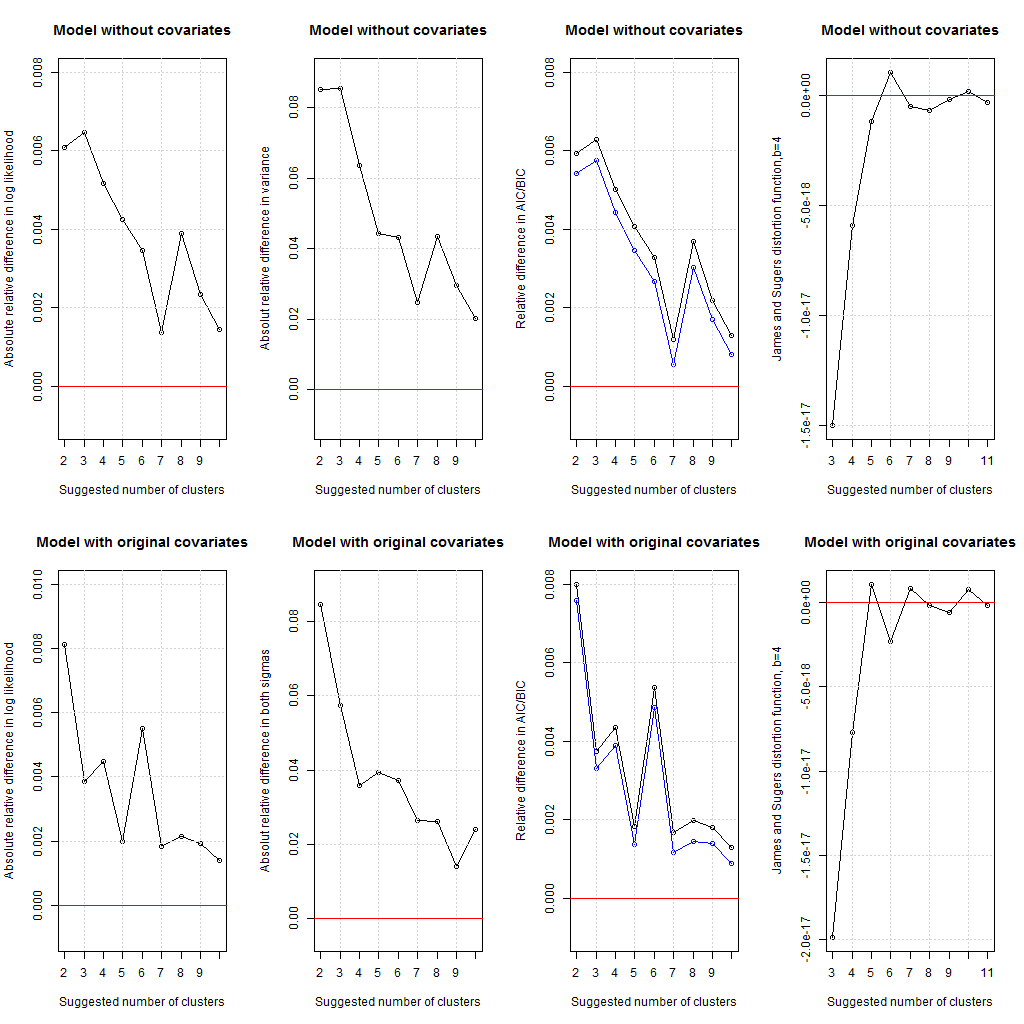}\caption{Deciding on the number of clusters for the Kassj\"o data changing from 2-11 clusters using four different measures. From left to right, the absolute relative difference in log likelihood, the absolut difference in the variance measure, the AIC and BIC measures and James and Sugars distortion measure, (with $b=4$), analyzed without covariates, upper row, and with original covariates, lower row.}
\label{reldiff}%
\end{figure}
\begin{figure}[ptb]
\centering
\textbf{Analysis with three original covariates}\par\medskip
\includegraphics[
height=6in,
width=\textwidth
]{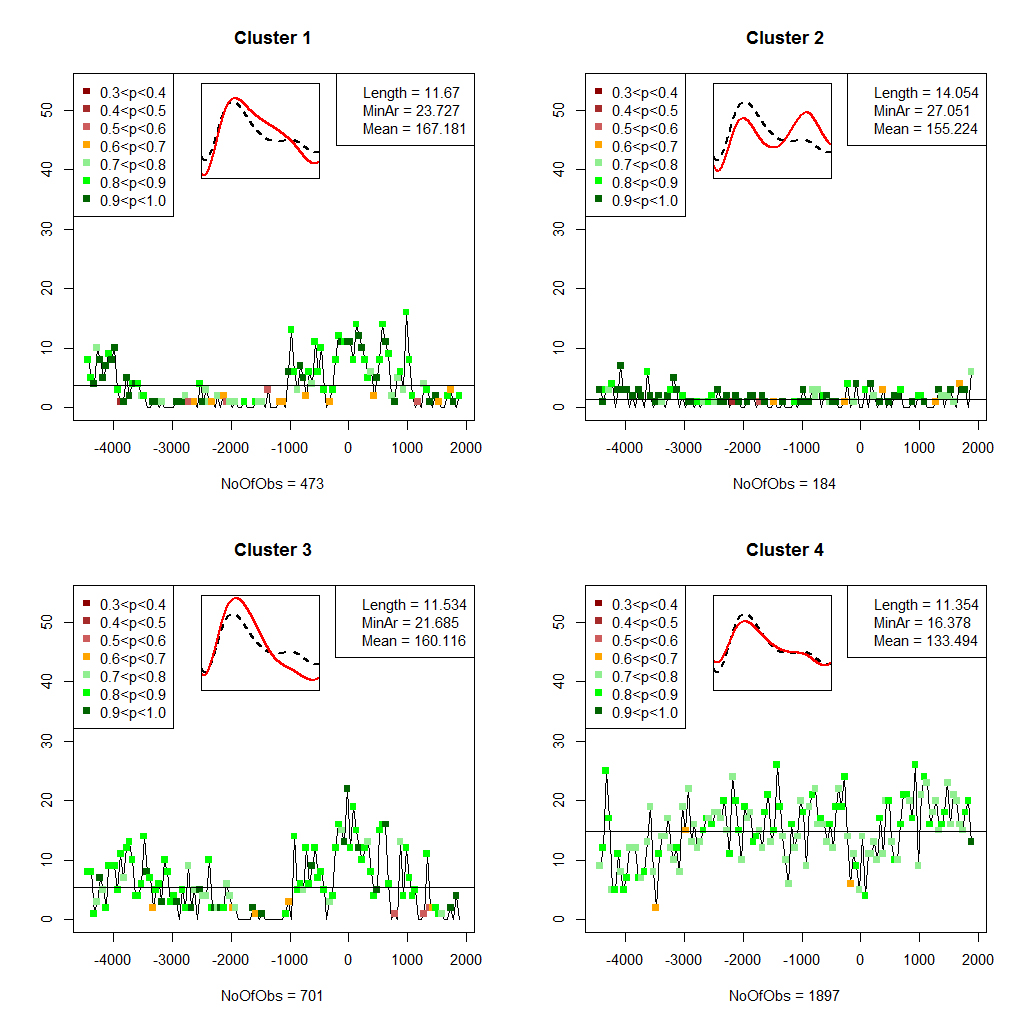}\caption{Dynamics of four of the seven clusters given by the frequencies of the different cluster types within 50-year periods (non-overlapping bins) starting from 1900 and going backwards. The profiles varies from sharp peak to flat peak and also a double peak. Posterior probabilities (means of 50 years as colored dots) are also given to indicate how uncertain the cluster frequencies are. Within each cluster the mean values is given for the included covariates in the analysis.}
\label{fourcluster}%
\end{figure}
\begin{figure}[ptb]
\centering
\textbf{Analysis with three original covariates}\par\medskip
\includegraphics[
height=6in,
width=\textwidth
]{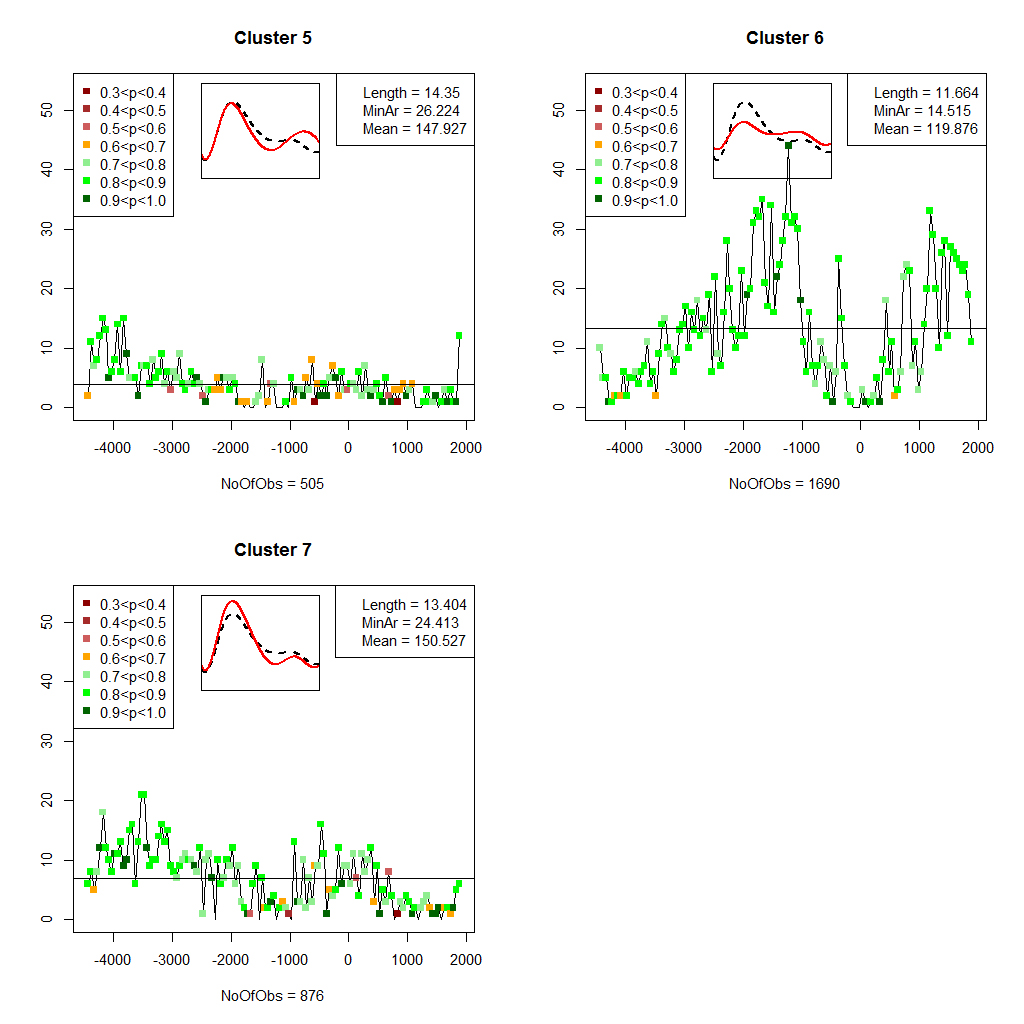}\caption{Dynamics of three of the seven clusters given by the frequencies of the different cluster types within 50-year periods (non-overlapping bins) starting from 1900 and going backwards. The profiles varies from sharp peak to flat peak and also a double peak. Posterior probabilities (means of 50 years as colored dots) are also given to indicate how uncertain the cluster frequencies are. Within each cluster the mean values is given for the included covariates in the analysis.}
\label{threecluster}%
\end{figure}
\begin{figure}[ptb]
\centering
\textbf{Analysis without covariates}\par\medskip
\includegraphics[
height=6in,
width=\textwidth
]{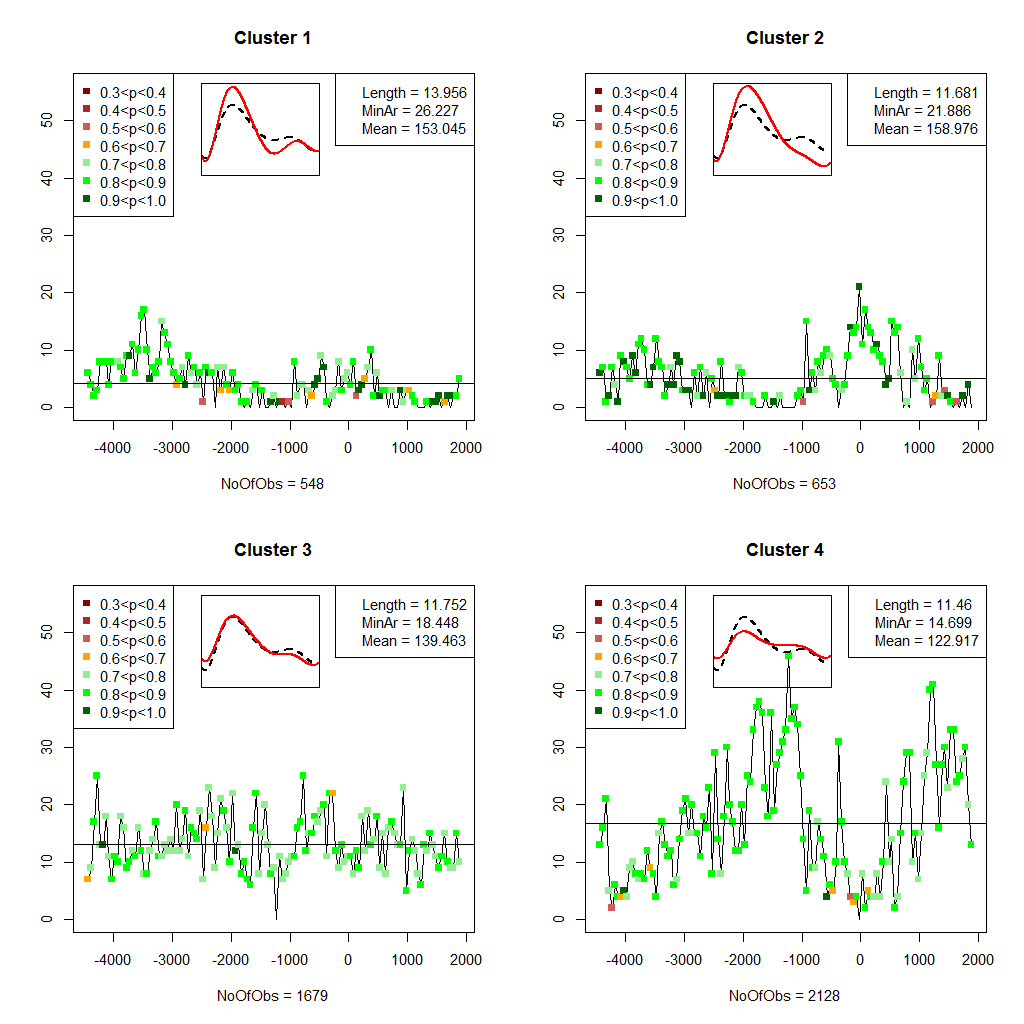}\caption{Dynamics of four of the seven clusters given by the frequencies of the different cluster types within 50-year periods (non-overlapping bins) starting from 1900 and going backwards. The profiles varies from sharp peak to flat peak and also a double peak. Posterior probabilities (means of 50 years as colored dots) are also given to indicate how uncertain the cluster frequencies are. The mean values, within each cluster is given for the covariates (not included in the analysis).}
\label{fourcluster.nocov}%
\end{figure}
\begin{figure}[ptb]
\centering
\textbf{Analysis without covariates}\par\medskip
\includegraphics[
height=6in,
width=\textwidth
]{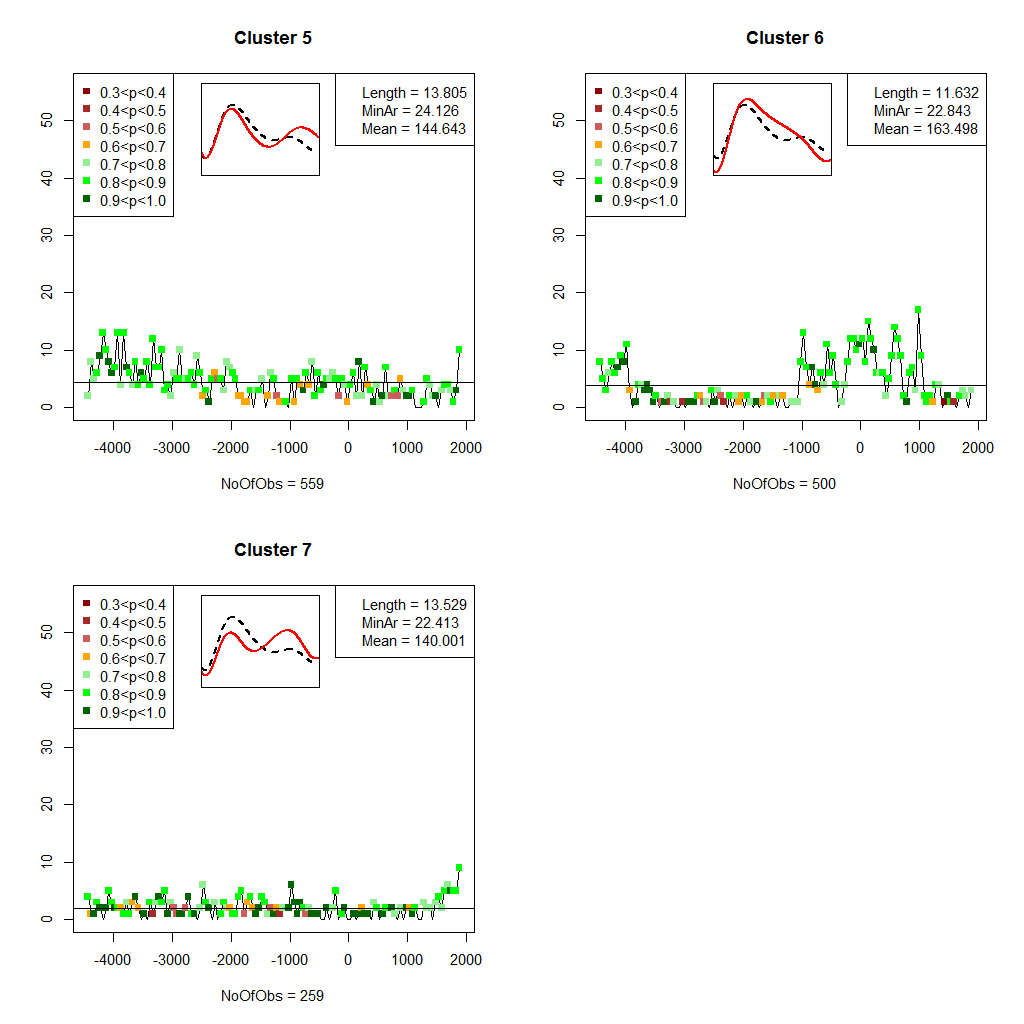}\caption{Dynamics of three of the seven clusters given by the frequencies of the different cluster types within 50-year periods (non-overlapping bins) starting from 1900 and going backwards. The profiles varies from sharp peak to flat peak and also a double peak. Posterior probabilities (means of 50 years as colored dots) are also given to indicate how uncertain the cluster frequencies are. The mean values, within each cluster is given for the covariates (not included in the analysis). }
\label{threecluster.nocov}%
\end{figure}
\begin{figure}[ptb]
\centering
\textbf{Illustration of posterior probabilities for year -3633}\par\medskip
\includegraphics[
height=6in,
width=\textwidth
]{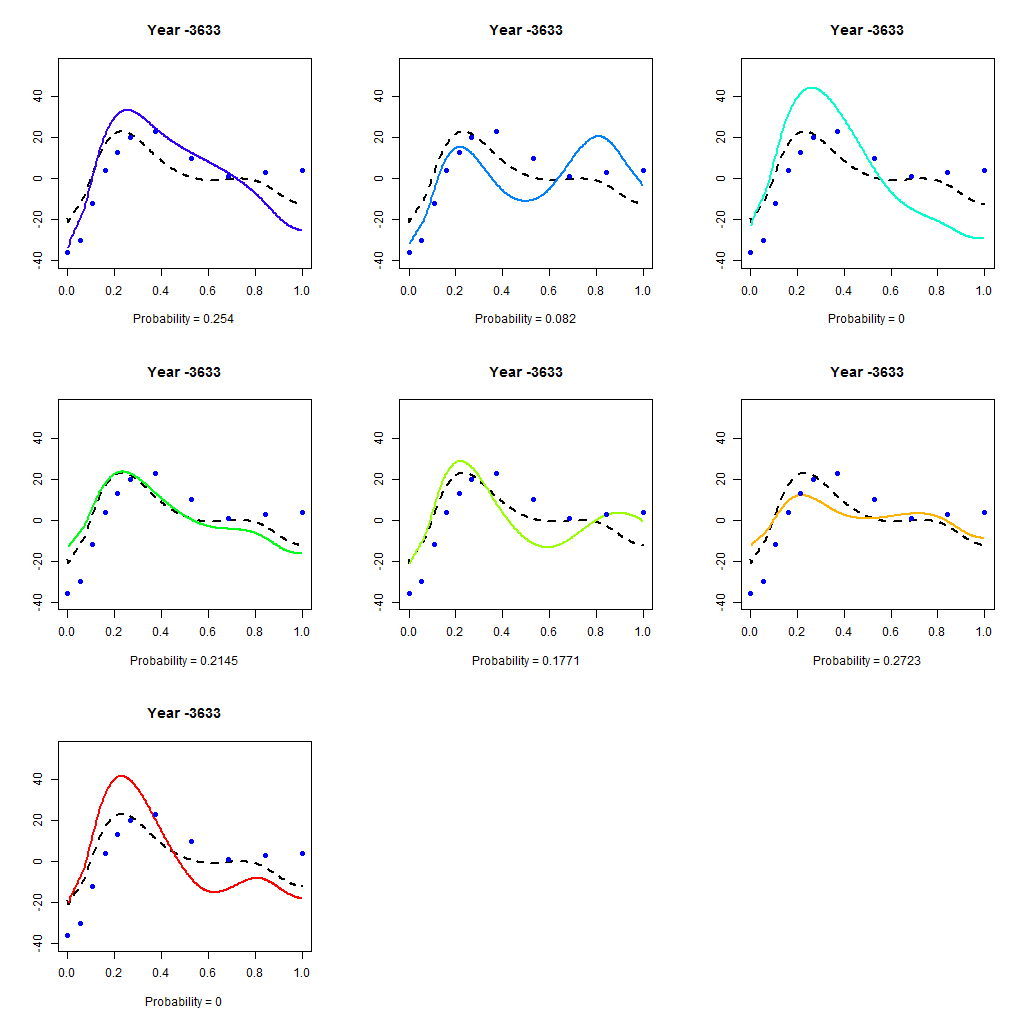}\caption{The year -3633 with its posterior probabilities for cluster belongings in the seven clusters. The solid colored curve is the mean cluster curve, the dashed line is the predicted curve within each cluster and the dots are the actual observations.}
\label{covill}%
\end{figure}
\begin{figure}[ptb]
\centering
\textbf{Illustration of posterior probabilities for year -668}\par\medskip
\includegraphics[
height=6in,
width=\textwidth
]{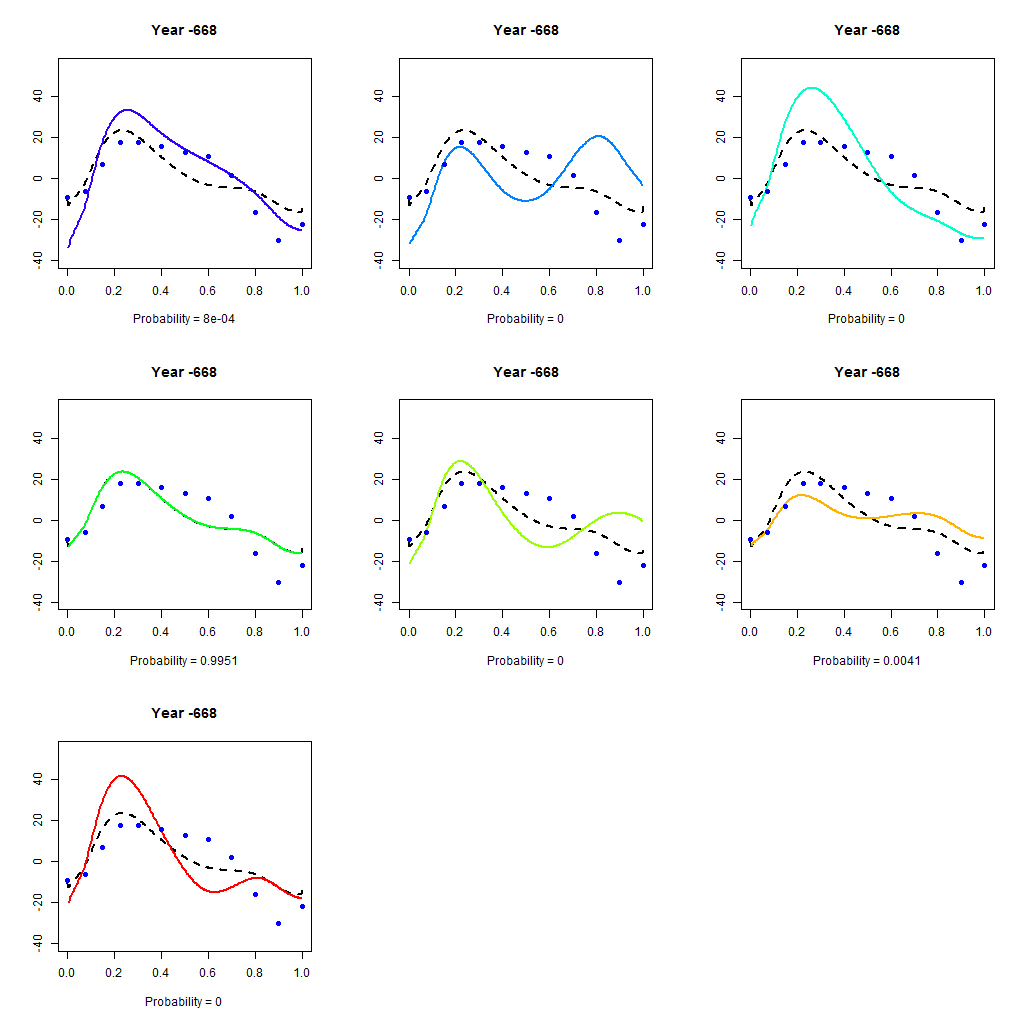}\caption{The year -668 with its posterior probabilities for cluster belongings in the seven clusters. The solid colored curve is the mean cluster curve, the dashed line is the predicted curve within each cluster and the dots are the actual observations.}
\label{covill2}%
\end{figure}
\begin{figure}[ptb]
\centering
\textbf{Boxplots of the three covariates}\par\medskip
\includegraphics[
height=6in,
width=\textwidth
]{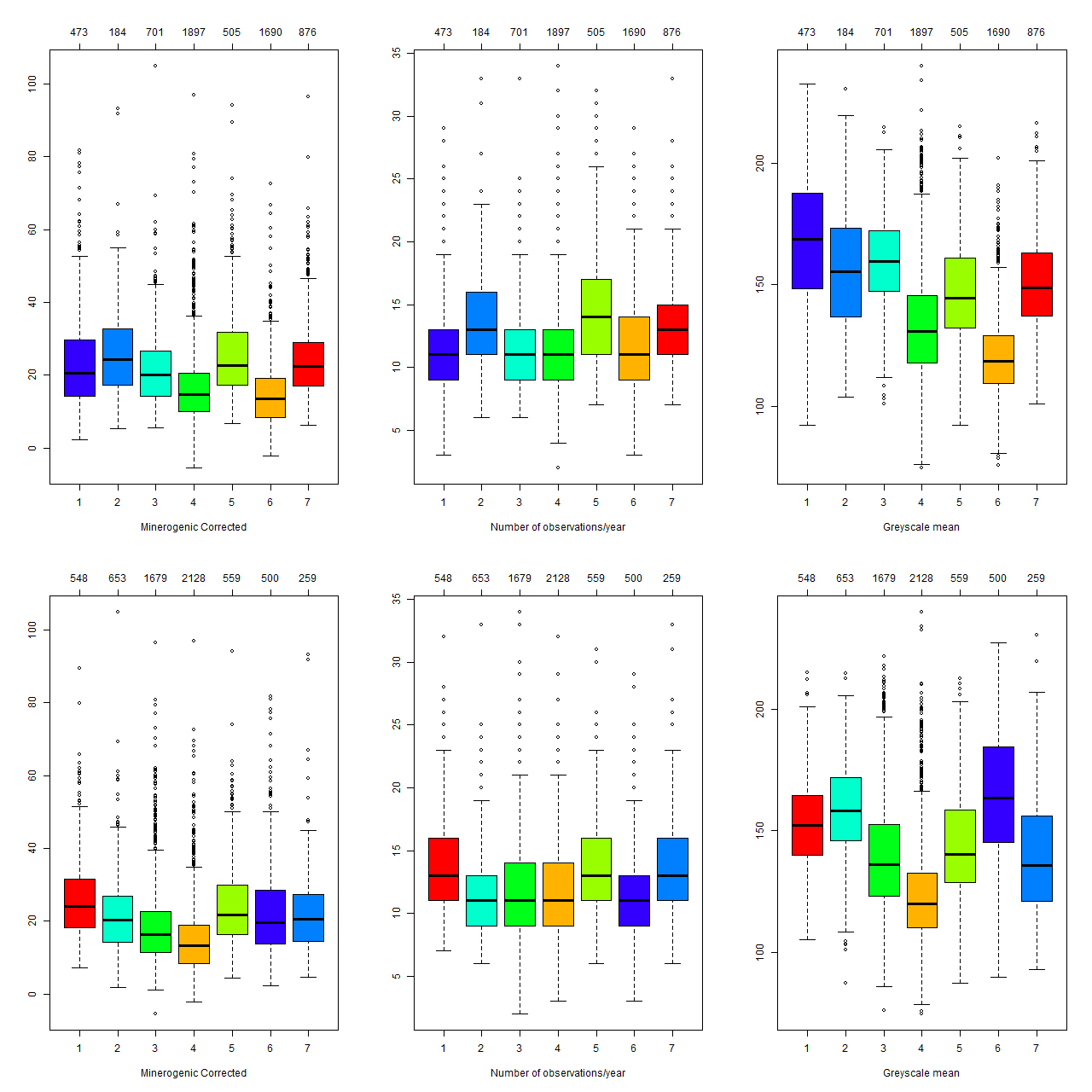}\caption{Boxplots of the covariates in the analysis for the 7 clusters. Upper row is when covariates is used in the modelling and the lower row is when no covariates was used in the modelling.}
\label{boxplots}%
\end{figure}
\begin{figure}[ptb]
\centering
\textbf{Covariance matrices, $\Delta_k$, k=1,...,7}\par\medskip
\includegraphics[
height=6in,
width=\textwidth
]{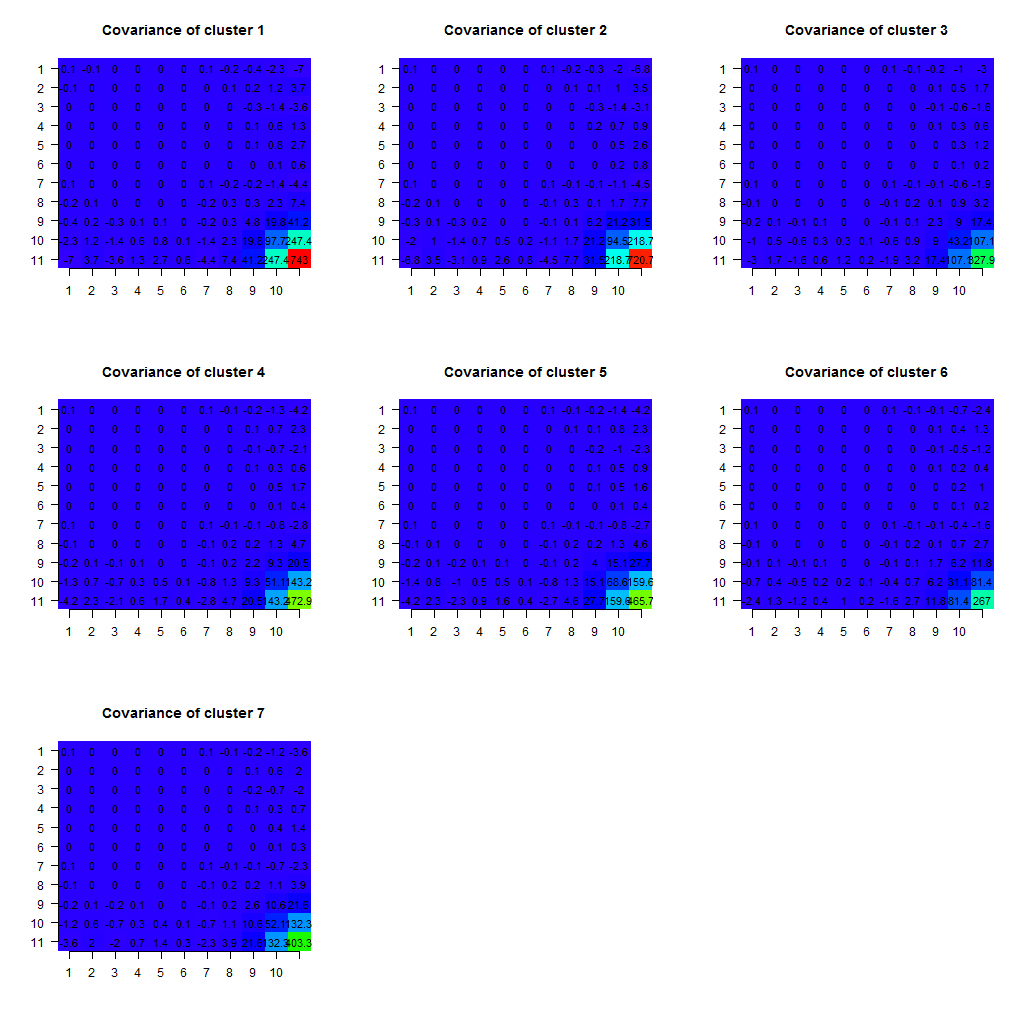}\caption{The Covariance matrices $\Delta_k$ for the 7 clusters illustrated by coloring the values in them. Red is the largest value and blue is the smallest value. Number 1--8 are the spline coefficients and 9 is MinAR, 10 is varve width and 11 is mean grey scale values.}
\label{covar}%
\end{figure}
\begin{figure}[ptb]
\centering
\textbf{Correlation matrices of $\Delta_k$, k=1,...,7}\par\medskip
\includegraphics[
height=6in,
width=\textwidth
]{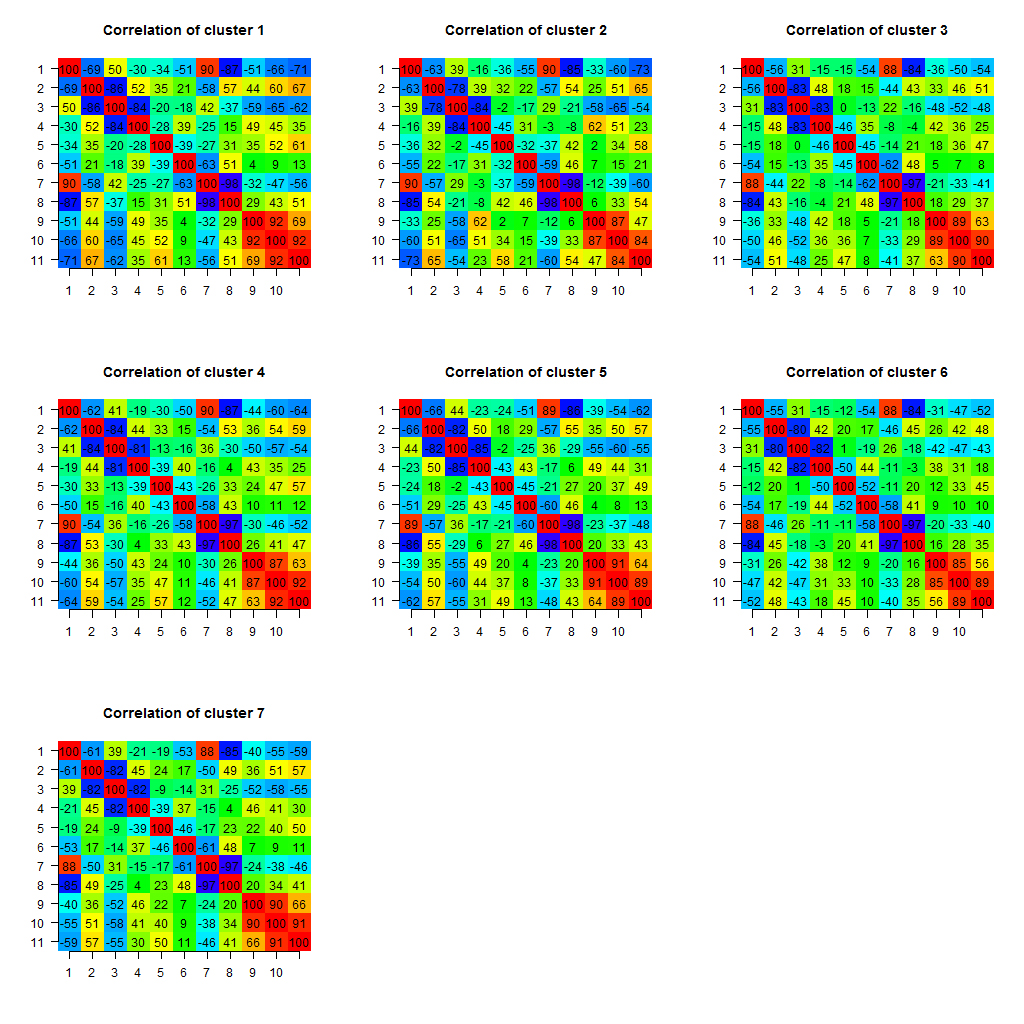}\caption{The Correlation matrices from the covariance matrices $\Delta_k$, for the 7 clusters illustrated by coloring the values in them. Red is the largest (positive) value and blue is the smallest (negative) value. Number 1--8 are the spline coefficients and 9 is MinAR, 10 is varve width and 11 is mean grey scale values.}
\label{corell}%
\end{figure}
\begin{figure}[ptb]
\centering
\textbf{Different basis representations}\par\medskip
\includegraphics[
height=6in,
width=\textwidth
]{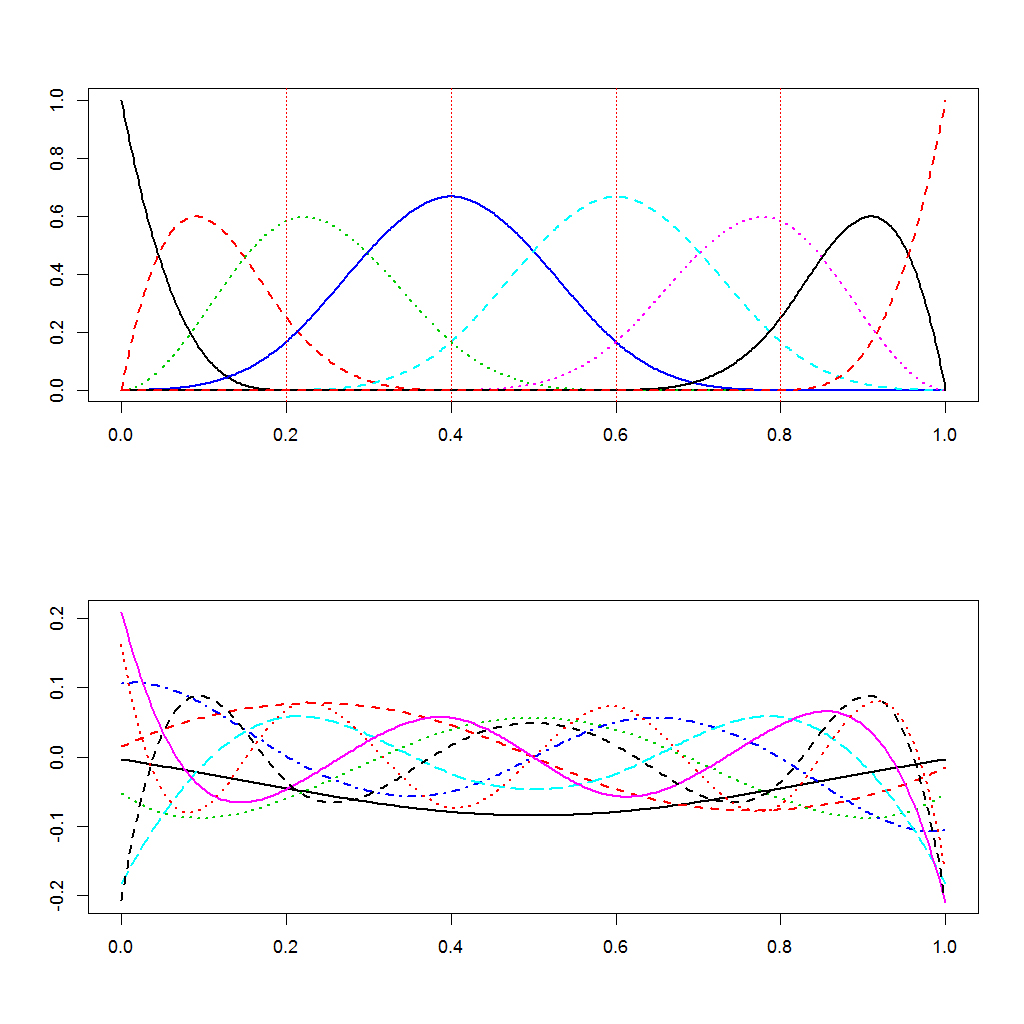}\caption{The picture shows what happens with the when shifting basis from cubic $b$-splines to the $\boldsymbol{U}$-matrix in the single value decomposition. Here 8 basis function are used.}
\label{basis}%
\end{figure}
\newpage
\clearpage
\bibliographystyle{chicago}
\bibliography{./referenser-v1}
\end{document}